\documentclass{iopart}
\usepackage{graphicx}
\usepackage{bm}
\usepackage{amsgen,amssymb,amsfonts,amsbsy}

\newcommand{\tend}{t_{\rm end}}
\newcommand{\Mfinal}{M_{\rm final}}
\newcommand{\Afinal}{A_{\rm f}}
\newcommand{\gH}{g_H}

\begin{document}

\title{Revisiting Event Horizon Finders}

\author{Michael I Cohen${}^1$, Harald P Pfeiffer${}^1$, Mark A Scheel${}^1$}

\address{${}^1$Theoretical Astrophysics, California Institute of
Technology, Pasadena, California 91125}

\date{\today}

\begin{abstract}
Event horizons are the defining physical features of black hole
spacetimes, and are of considerable interest in studying black hole
dynamics.  Here, we reconsider three techniques to find event horizons
in numerical spacetimes: integrating geodesics, integrating a surface,
and integrating a level-set of surfaces over a volume.  We implement
the first two techniques and find that straightforward integration of
geodesics backward in time is most robust.  We find that the
exponential rate of approach of a null surface towards the event
horizon of a spinning black hole equals the surface gravity of the
black hole.  In head-on mergers we are able to track quasi-normal
ringing of the merged black hole through seven oscillations, covering
a dynamic range of about $10^5$.  Both at late times (when the final
black hole has settled down) and at early times (before the merger),
the apparent horizon is found to be an excellent approximation of the
event horizon.  In the head-on binary black hole merger, only {\em
  some} of the future null generators of the horizon are found to
start from past null infinity; the others approach the event horizons
of the individual black holes at times far before merger.

\end{abstract}



\section{Introduction}
\label{s:Introduction}

The two body problem in general relativity has been the focus of
extensive work for many years and, because there is no analytic
solution, it must be solved numerically.  Binary black hole mergers
are expected to be one of the most astrophysically common sources of
gravitational radiation for detectors such as
LIGO~\cite{Barish:1999,Waldman:2006}.  Recent advances in simulating
binary black hole mergers include the development of the generalized
harmonic evolution system~\cite{Pretorius2006} and the moving
punctures technique~\cite{Campanelli2006a,Baker2006a}.  In the last
several years the field has reached a stage where binary black hole
simulations are becoming routine.  Numerical simulations have been
remarkably successful in expanding our understanding of binary black
holes, but challenges remain.

One particular challenge is to be able to more accurately locate the
holes during the merger.  There are two useful concepts to describe
the location of black holes in a spacetime, {\em apparent horizons}
(AH) and {\em event horizons} (EH).  An EH is the true surface of a
black hole: it is defined as the boundary of the region of the
spacetime that is causally connected to future null infinity.  Because
the definition of the EH involves global properties of the spacetime,
one must know the {\em full} future evolution of the spacetime before
the EH can be determined exactly.  This difficulty has led researchers
to instead identify black holes with apparent horizons, which are
defined in terms of the expansion of null congruences.
\footnote{More precisely, we define AH as the outermost marginally
  outer-trapped surface, where an outer-trapped surface is a
  topological 2-sphere with zero expansion along outgoing null
  normals.} Indeed, AH finders are highly developed and have been the
subject of extensive work (see, e.g. the review~\cite{Thornburg:2006})
Unlike an EH, an AH can be located from data on a single spacelike
hypersurface, i.e. on each timestep of a numerical evolution, without
knowing the future evolution of the spacetime.  The AH is often an
effective substitute for the EH for several reasons. First, according
to the cosmic censorship conjecture, if an AH is present, it must be
surrounded by an EH.  Second, if an AH is present on a spacelike
hypersurface through a stationary spacetime, it coincides with the EH.
Finally, in numerical simulations, apparent horizons generally show
behaviour attributed to event horizons: For instance, the area of the
AH typically does not decrease and it is usually almost constant
whenever the spacetime is only mildly dynamic.  In fact, apparent
horizons have motivated the development of ``isolated'' and
``dynamical'' horizons (see~\cite{Ashtekar-Krishnan:2004} for a
review).  These surfaces satisfy analogues of the laws of black hole
thermodynamics, although they are defined quasi-locally, rather than
globally.

However, using the AH to locate the holes is not always
appropriate. For instance, the AH is slicing dependent, while the EH
is not.  Indeed, the Schwarzschild spacetime can be sliced in such a
way that no AH exists~\cite{Wald91}.  Furthermore, even on slicings on
which an AH is present, there are few precise mathematical statements
about how ``close'' AH and EH are.  Finally, AH and EH behave
qualitatively differently during a black hole merger: The EH
\footnote{More precisely, the 2-surface formed by the intersection of
  the spatial slice and the EH} around each black hole expands
continuously until the two components of the EH join into one, whereas
a common apparent horizon appears discontinuously quite some time
after the EHs have merged. The common AH encompasses the two
individual AHs, which continue to exist as surfaces of zero outgoing
null expansion for some time after the merger.

Early EH finders~\cite{Shapiro-Teukolsky:1980,Hughes94} followed null
geodesics forward in time and determined whether or not each geodesic
eventually escapes to infinity.  Following geodesics forward in time
is unstable in that slightly perturbed geodesics will diverge from the
EH and either escape to infinity or fall into the singularity.
Furthermore, a large number of geodesics with different directions
must be sampled at each point and at each time step to determine if
one of these succeeds in escaping to infinity~\cite{Hughes94}.
To reduce the number of sampling points,
the EH search in~\cite{Hughes94}
was performed on a series of time slices proceeding backward from late to early
times; to find the EH on each time slice,
they integrated geodesics forward in time,
using the already-located EH at the later time as an initial guess.

Since outgoing null geodesics diverge from the event horizon when
going forward in time, when going {\em backward} in time they will
converge onto the event horizon~\cite{Libson95a,Libson96}.  All recent
EH finders use this observation, and follow null geodesics or null
surfaces backward in
time~\cite{Thornburg:2006,Libson95a,Libson96,matzner_etal95,WalkerThesis:1998,MassoEtAl:1999,CavenyThesis:2002,Caveny2003,Caveny2003a,Diener:2003}.

Several algorithms have been developed to follow null geodesics
backward in time.  These can be divided into three types, which we
shall refer to as the ``geodesic method'', the ``surface method'' and
the ``level-set method''. The geodesic method works by simply
integrating the geodesic equation, as done by Libson
et. al.~\cite{Libson96}.  Libson et. al. express concerns that the
geodesic method may be susceptible to tangential ``drifting'' of the
geodesics.  However, this is not evident when the method is applied to
the science applications in that paper, nor do we find tangential
drifting in our simulations.  To avoid any issues with drifting,
Libson et al.  introduced the surface method: a complete null surface
(rather than individual geodesics) is evolved backward in time.
In~\cite{Libson95a,Libson96} this surface was parameterized based on
axisymmetry (although the parameterization of
\cite{Libson95a,Libson96} cannot handle {\em generic} axisymmetric
situation, cf. Section~\ref{ss:SurfaceMethod} below), and many
interesting results on the structure of caustics and the geometry of
the horizon for axisymmetric spacetimes were obtained in
\cite{matzner_etal95,MassoEtAl:1999}.  Diener~\cite{Diener:2003} and
Caveny et. al~\cite{CavenyThesis:2002,Caveny2003,Caveny2003a}
independently introduced the level-set method by recasting the surface
method in a way that does not assume symmetry: rather than evolving a
single 2-D surface, they evolve a volume-filling series of surfaces
given as the level-sets of a spacetime function $f(t,x^i)$.  To avoid
exponentially steepening gradients of $f$, Caveny et al introduce an
artificial diffusive term, whereas Diener reinitializes $f$ whenever
necessary.

This paper re-examines these techniques for event horizon finding in
the context of the Caltech/Cornell Spectral Einstein Code (SpEC),
which provides an infrastructure for highly accurate simulations of
Einstein's equations for single and binary black holes.  Recent work
includes highly accurate computations of gravitational waveforms from
inspiraling
binaries~\cite{Pfeiffer-Brown-etal:2007,Boyle2007,Boyle2008a}.  The
availability of high accuracy binary evolutions motivates the
development of very precise event horizon finding techniques in order
to extract all possible physics from these simulations.  Therefore,
this paper reconsiders the three techniques mentioned above in the
context of general binary black hole mergers without any symmetries.

We implement the geodesic method, and generalize the surface method to
arbitrary situations without symmetries.  Both methods are then
applied to single Kerr black holes, and a head-on binary black hole
merger.  In both cases, the geodesic method is found to be more
robust.  We encounter two fundamental problems with the level-set
method, and therefore halted our efforts to implement it in SpEC.

This paper is organized as follows: In Section~\ref{s:Methods}, we
explain the three methods in more detail and give details of our
numerical implementation.  Section~\ref{s:KerrApplication} presents
results for a single Kerr black hole, and in
Section~\ref{s:HeadonMergerApplication}, we apply the techniques to a
head-on BBH merger, where we extract ringdown behaviour and the
behaviour of the individual event horizons before merger.  We close
with a conclusion in Section~\ref{s:Conclusion}.

\section{Methods}
\label{s:Methods}

All EH-finding techniques considered here proceed backward in time and
must therefore be performed after the numerical evolution of the
spacetime has been completed.  We assume that we have access to the
spacetime metric in a 3+1 decomposition
\begin{equation}
ds^2 = -N^2dt^2 + \gamma_{ij}(dx^i + \beta^idt)(dx^j +
\beta^jdt) \label{e:3+1_decomp},
\end{equation}
where $N$ is the lapse, $\beta^i$ is the shift, and $\gamma_{ij}$ is
the 3-metric on the slice. Latin indices $i,j,\ldots=1,2,3$ denote
spatial dimensions; below we will use Greek indices to denote
spacetime dimensions, $\alpha, \beta \ldots =0,1,2,3$.  The time $t$
in (\ref{e:3+1_decomp}) represents the coordinate time of the
numerical evolution.  Typically, the metric data $\gamma_{ij}$,
$\beta^i$, and $N$ are available at discrete times and at discrete
spatial grid points.  Evaluating the values of the metric components
elsewhere requires interpolation.

\begin{figure}
\centerline{
  \includegraphics[width=.9\textwidth]{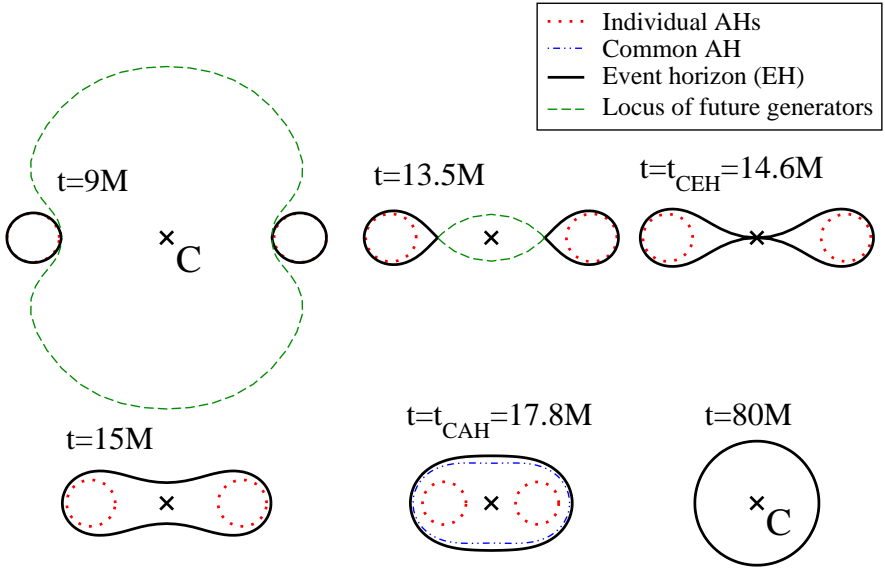} }
\caption{Cross-sections through event and apparent horizons during a
  BBH merger.  Before the merger, $t<t_{\rm CEH}$, the surface
  includes the set of generators that will merge onto the event
  horizons through the cusps in the individual event horizons (green
  dashed curves).  The point C is the point of symmetry for the
  head-on merger, which will be used in Section~\ref{ss:LevelSet}.
\label{fig:MultipleCross-sections}}
\end{figure}

A black-hole merger exhibits several characteristic features of
relevance to EH finders, as illustrated in
Figure~\ref{fig:MultipleCross-sections}\footnote{While
  Figure~\ref{fig:MultipleCross-sections} is meant as an illustration,
  it presents actual data from the head-on binary black hole merger
  discussed in Section~\ref{s:HeadonMergerApplication}.  The time
  given at each frame of Figure~\ref{fig:MultipleCross-sections} will
  aid in the discussion in
  Section~\ref{s:HeadonMergerApplication}.}\,(cf.~\cite{Libson96,Caveny2003a,MassoEtAl:1999}).
At times sufficiently far prior to merger, the EH and AH are expected
to coincide closely (and indeed, we confirm this below for our
simulation).  The green dashed curves in
Figure~\ref{fig:MultipleCross-sections} represent future generators of
the event horizon, i.e. null geodesics that will merge onto the event
horizon through cusps in the individual event horizons. These cusps
are clearly visible at time $t=13.5M$ where the individual EHs have
diverged significantly from their respective AHs.  At $t_{\rm CEH} =
14.6M$ the two previously disjoint components of the event horizon
join.  We shall refer to this time $t_{\rm CEH}$ as {\em the merger}
of the black hole binary.  After the merger, the event horizon of the
merged black hole can be seen relaxing towards its final
time-independent shape.  The common apparent horizon appears at
$t_{\rm CAH} = 17.8M$, and approaches the event horizon as the
evolution proceeds; at $t=80M$, the AH coincides almost exactly with
the event horizon.

\subsection{Geodesic Method}
\label{ss:GeodesicMethod}

The most straightforward way to follow light rays is to simply
integrate the geodesic
equation~\cite{Shapiro-Teukolsky:1980,Hughes94,Libson95a,Libson96},
\begin{equation}
\frac{d^{2}q^{\mu}}{d\lambda^{2}} + \Gamma^{\mu}_{\alpha\beta}
\frac{dq^{\alpha}}{d\lambda} \frac{dq^{\beta}}{d\lambda} =
0, \label{e:geodesic}
\end{equation}
where $q^\mu=q^\mu(\lambda)$ is the position of the photon on the
geodesic, parameterized by an affine parameter $\lambda$, and
$\Gamma^{\mu}_{\alpha\beta}$ are the spacetime Christoffel symbols.

Since we have access to our spacetime as a function of the evolution
time coordinate $t$, it is convenient to rewrite (\ref{e:geodesic}),
replacing $\lambda$ by $t$ along the geodesic.  Writing $\dot q^\mu =
dq^{\mu}/dt$, and $a = d\lambda/dt$, we find:
\begin{eqnarray}
\frac{dq^{\mu}}{d\lambda} &=& \frac{1}{a}\dot q^\mu, 
\\ \frac{d^2q^{\mu}}{d\lambda^2} &=& \frac{1}{a^2}\ddot q^\mu -
\frac{\dot a}{a^3} \dot q^\mu. 
\end{eqnarray}
Substituting into the geodesic equation we get 
\begin{equation}
\ddot q^\mu = \frac{\dot a}{a}\dot q^\mu - \Gamma^\mu_{\alpha\beta}
\dot q^\alpha \dot q^\beta. \label{e:qdotprelimequation}
\end{equation}
The quantity $a$ is determined by the requirement that $q^0 = t$,
i.e. that at parameter value $t$ along the geodesic, the geodesic is
on the corresponding $t=\mbox{const}$ hypersurface of the evolution.
This implies $\dot q^\mu = [1,\dot q^i]$ and $\ddot q^\mu = [0,\ddot
q^i]$. Setting $\mu=0$ in (\ref{e:qdotprelimequation}) gives
$\frac{\dot a}{a} = \Gamma^0_{\alpha \beta} \dot q^\alpha \dot
q^\beta$. The spatial components of (\ref{e:qdotprelimequation}) are
the desired evolution equation for the spatial coordinates as a
function of coordinate time,
\begin{equation}
\ddot q^i = \Gamma^0_{\alpha\beta} \dot q^\alpha \dot q^\beta \dot q^i
- \Gamma^i_{\alpha\beta} \dot q^\alpha \dot q^\beta.
\end{equation}
We convert this set of ordinary differential equations to first order
form by defining $p^i \equiv \dot q^i$, which gives 
\numparts
\begin{eqnarray}
\dot q^i =
p^i, \label{e:GeodesicMethodEvolutionEquation2} \\
\dot p^i = \Gamma^0_{\alpha\beta} p^\alpha p^\beta p^i -
\Gamma^i_{\alpha\beta} p^\alpha
p^\beta, \label{e:GeodesicMethodEvolutionEquation1}
\end{eqnarray}
\endnumparts
facilitates the use of standard ODE integrators like Runge-Kutta
methods~\cite{numrec_c,ButcherNumericalMethods}.

While integrating geodesics is not new~\cite{Hughes94,Libson96},
re-expressing the geodesic equation in terms of coordinate time seems
to be new.  It appears that the primary reason this technique has been
phased out in favour of the two techniques described below is the
concern that, in a full 3D implementation, slight tangential
velocities may be imparted to the outgoing null geodesics through
numerical inaccuracies, and that this tangential drift of geodesics
could result in unphysical caustics.  These concerns are discussed in
detail in~\cite{Libson96}, where the idea of representing the whole
surface, rather than individual geodesics, was introduced.  This was
justified on the basis that for a surface, tangential drift is
irrelevant.  However, while it is possible that tangential drift can
be significant for very coarse, low-resolution simulations, we see no
evidence that tangential drift affects our numerical tests of the
geodesic method.

We finally like to point out that if one evolves $p_i=g_{i\mu}p^\mu$
instead of $p^i$ (cf. \cite{Hughes94}), then the evolution equations
depend only on spatial derivatives of the spacetime metric.  Evolving
$p_i$ therefore results in computational savings, because the time
derivatives of the metric need not be stored or interpolated. This
will be investigated in a future work.

\subsection{Surface Method}
\label{ss:SurfaceMethod}

\begin{figure}
\centerline{\includegraphics[width=.90\textwidth]{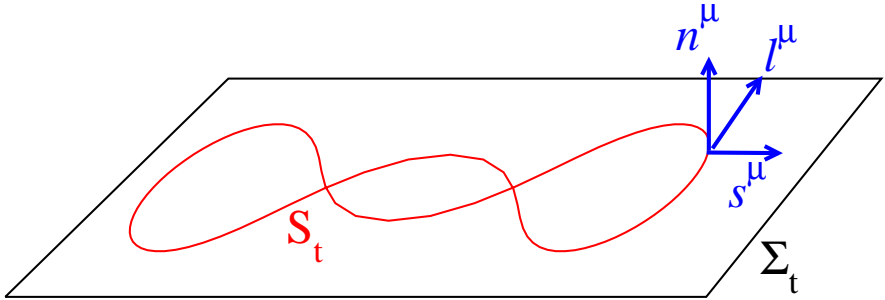}}
\caption{A slice of the event horizon $(\mathcal{S}_t)$, produced by
  the intersection of a spatial hypersurface $\Sigma_t$ with the world
  tube of the event horizon $\mathcal{N}$, showing $n^\mu$ (the
  timelike normal to $\Sigma_t$), $s^\mu$ (the spatial normal of
  $\mathcal{S}_t$), and $\ell^\mu$ (the null normal to the EH
  world-tube $\mathcal{N}$).
\label{fig:Cross-sectionFig}}
\end{figure}

The idea of the surface method dates back to Libson et
al.~\cite{Libson96}, who used it in axisymmetry.  The goal is to
evolve a 2-dimensional surface $\mathcal{S}_t$ backward in time such
that it traces out a null hypersurface $\mathcal{N}$.  The time
coordinate $t$ is inherited from the black hole simulation for which
event horizons are to be determined, i.e. $\mathcal{S}_t$ is the
intersection of $\mathcal{N}$ with the spatial hypersurfaces
$\Sigma_t$ of the evolution, as indicated in
Figure~\ref{fig:Cross-sectionFig}.  Before the black hole merger,
$t<t_{\rm CEH}$, the surface $\mathcal{S}_t$ consists not only of the
two disjoint parts of the event horizon, but also includes the future
generators, which are indicated by the green dashed curves in
Figure~\ref{fig:MultipleCross-sections}.  The union of these three
components is a smooth self-intersecting surface with the topology of
a sphere (as suggested by Kip Thorne\cite{Libson95a,Libson96}).

Let us first consider how to represent the surface to be evolved.
Apparent horizon finders often parameterize a surface by giving the
radius, relative to a fixed point, as a function of angular
coordinates, i.e. $r=f(\theta,\phi)$.  Such a star-shaped surface is
insufficient here, because the surface will be self-intersecting for
$t<t_{\rm CEH}$ and will cease to be star-shaped even before then (see
Fig. 1.1 of~\cite{Thornburg:2006}).  The axisymmetric EH finder
presented in~\cite{Libson96} parameterized the surface by $\rho=s(z,
t)$, where $z$ is a coordinate along the axis of symmetry, and $\rho$
is the cylindrical radius.  This allows some mild form of
self-intersection, like, for instance, the $t=13.5M$ snapshot in
Figure~\ref{fig:MultipleCross-sections}.  However, at earlier times,
the locus of future null generators of the horizon ``bulges outward''
and becomes multivalued when considered as a function of $z$,
cf. $t=9M$ in Figure~\ref{fig:MultipleCross-sections}.  In this case,
the parameterization of~\cite{Libson96} fails even for an axisymmetric
configuration.  In this paper, we use a parametric representation of
$\mathcal{S}_t$, i.e. $r^i=\left.r^i(t, u, v)\right|_{t}$.  The full
3-dimensional null hypersurface ${\cal N}$ being constructed is
represented as a 3-parameter surface in spacetime:
\begin{equation}
r^\mu(t,u,v) =
\big[t,r^i(t,u,v)\big].\label{eq:Surface-Parameterization}
\end{equation}

We wish to find an equation that will allow us to evolve
$\mathcal{S}_t$ in such a way as to trace out the null 3-surface
$\mathcal{N}$.  Further, we would like this equation to have the
property that for fixed $(u_0,v_0)$, the curve $r^\mu(t,u_0,v_0)$
traces out a null geodesic.  This allows us to directly compare the
surface obtained by the surface method to the surface obtained for
equivalent initial conditions by the geodesic method.

For the curve $\left.r^\mu(t, u, v)\right|_{u,v}$ to be null, its
tangent $\partial r^\mu(t,u,v)/\partial t$ must be outgoing and null,
i.e.
\begin{equation}
\frac{\partial r^\mu}{\partial t} =
\ell^\mu, \label{e:surfaceevolutionequation}
\end{equation}
where $\ell^\mu$ is a null normal to $\mathcal{S}_t$.  $\ell^\mu$ can
be written as
\begin{equation}
\ell^\mu = c (n^\mu + s^\mu), \label{e:defn_of_l}
\end{equation}
where $n^\mu$ is the timelike unit normal to $\Sigma_t$, $s^\mu$ is
the spatial outward-pointing unit normal to $\mathcal{S}_t$
(cf. Figure~\ref{fig:Cross-sectionFig}) and $c$ is an overall scaling.
Consistency of (\ref{eq:Surface-Parameterization}) and
(\ref{e:surfaceevolutionequation}) requires that $\ell^\mu$ is
normalized such that $\ell^t=1$.  To find the value of $c$ from the
condition $\ell^t=1$, first notice that from the 3+1 decomposition,
\begin{equation}
n^\mu = \frac{1}{N}\big[1,-\beta^i\big],
\end{equation}
where $N$ and $\beta^i$ are the lapse and shift fields.  Also,
since $s^\mu$ lies within the spatial slice $\Sigma_t$, we may write
$s^\mu = \big[0,s^i\big]$, so that (\ref{e:defn_of_l}) becomes
\begin{equation}
\ell^\mu = c \left[\frac{1}{N},s^i - \frac{1}{N} \beta^i\right].
\end{equation}
Thus $\ell^t=1$ implies $c = N$, and we can write our final evolution
equation for the spatial components of $r^i$,
\begin{equation}
\frac{\partial r^i}{\partial t} = Ns^i -
\beta^i. \label{e:SurfaceMethodEvolutionEquation}
\end{equation}

In order to find the unit normal $s^i$ to the spatial surface ${\cal
  S}_t$, we follow the standard procedure for a surface parameterized
as $r^i(u,v)$, i.e.
\numparts
\begin{eqnarray}
 \tilde{s}^i &=& \gamma^{il}\epsilon_{ljk} \frac{\partial
   r^j}{\partial u} \frac{\partial r^k}{\partial
   v}, \label{e:SurfaceSpatialDerivatives}\\ \rho &=&
 \sqrt{\gamma_{ij}\tilde{s}^i\tilde{s}^j},\\ s^i &=& \rho^{-1}
 \tilde{s}^i.\label{e:Signoflambda}
\end{eqnarray}
\endnumparts
where $\epsilon_{ljk}$ is the antisymmetric tensor and where we have
chosen the sign of the root such that $s^i$ points outward for a
right-handed choice of coordinates.

This evolution equation for the surface method
(\ref{e:SurfaceMethodEvolutionEquation}) is very different from the
evolution equations for the geodesic equation
(\ref{e:GeodesicMethodEvolutionEquation1})-(\ref{e:GeodesicMethodEvolutionEquation2}).  The surface method
does not require derivatives of the metric, but derivatives
$\partial_u r^i$, $\partial_v r^i$ along the surface; the geodesic
method, in contrast, requires derivatives of the metric, but treats
each geodesic completely independently.  Nevertheless, due to our
choice of evolution equation (\ref{e:surfaceevolutionequation}), each
point on the parameterization of the surface traces its own geodesic;
see \ref{App:SurfaceGeodesicProof} for a proof.

\subsection{Level-set Method}
\label{ss:LevelSet}

The level-set
method~\cite{Libson96,WalkerThesis:1998,MassoEtAl:1999,CavenyThesis:2002,Caveny2003,Caveny2003a,Diener:2003}
utilizes a function $f=f(t,x^i)$ defined on the full spacetime (or at
least, a region of spacetime covering the vicinity of the expected
location of the EH).  The function $f$ is determined such that
$f=\mbox{const}$ contours (i.e. level-sets) represent null surfaces
i.e. $g^{\alpha\beta}\partial_\alpha f\partial_\beta f = 0$.  In the 3+1 decomposition,
this becomes~\cite{Caveny2003,Caveny2003a,Diener:2003},
\begin{equation}
\partial_t f = \beta^i \partial_i f \mp N\sqrt{\gamma^{ij} \partial_i
  f\partial_j f}, \label{e:VolumeMethod}
\end{equation}
where the $\mp$ accommodates both ingoing and outgoing null surfaces,
with the minus sign being appropriate for outgoing null surfaces if
the gradient $\partial_i f$ is outward-pointing.

Libson et al~\cite{Libson96} had previously made use of
(\ref{e:VolumeMethod}), but parameterized the $f=0$ contour based on
axisymmetry.  The motivation of evolving~(\ref{e:VolumeMethod})
directly in the volume is to remove any assumptions of symmetries.

Unfortunately, 
when trying to implement the level-set method in SpEC, we encountered
two fundamental problems.  The first difficulty is related to the
characteristic speed of the level-set method.  Simply put, all
$f=\mbox{const}$ contours approach the event horizon, therefore new
contours need to be filled in at the boundaries of the region in which
$f$ is evolved (i.e. the outer boundary and possibly one or more inner
boundaries if black hole excision is employed).  To see this, note that the
characteristic speed of (\ref{e:VolumeMethod}) relative to a spatial
direction $\bar{n}_i$ is
\begin{equation}
v=N\bar{n}_i\frac{\partial^i f}{\sqrt{\gamma^{ij} \partial_i f
    \partial_j f}}-\bar{n}_i\beta^i,
\label{e:CharSpeed}
\end{equation}
where the sign of the first term depends on the gradient $\partial_i
f$ being outward pointing.  For most coordinate systems of interest,
lapse $N$ and shift $\beta^i$ behave such that $v>0$ at the outer
boundary and at any excision boundaries (if present).  When
integrating (\ref{e:VolumeMethod}) backward in time, well-posedness
requires boundary conditions at these boundaries.  Our preferred
numerical techniques are spectral methods because of their promise to
achieve exponential convergence for smooth problems.  Spectral methods
are very sensitive to the existence of an underlying well-posed
continuum problem and therefore require boundary conditions.
Unfortunately there is no particular physical reasoning to suggest a
choice of boundary condition.  While essentially any choice of
boundary condition that results in $f$ being continuous rendered our
spectral level-set implementation stable, and convergent to at least
first order, we have been unable to find a boundary condition that
ensures that $f$ remains smooth and thus leads to the desired
exponential convergence, not even in the single black hole case.  A
full finite-difference evolution of $f$ would be less sensitive to the
lack of proper boundary conditions (see~\cite{Diener:2003}), but would
be much slower for finding an EH in spectral-code metric data (due to
interpolations from the spectral to the finite-difference grid) and
much less accurate.

The second fundamental difficulty lies in singular behaviour of the
function $f$ in certain cases.  Let us consider an equal-mass head-on
merger as depicted in Figure~\ref{fig:MultipleCross-sections}.  Assume
$f$ to be smooth, and let us focus on the value of $f$ at the point of
symmetry, marked with C in Figure~\ref{fig:MultipleCross-sections}.
We assume that $\partial_i f$ is outward-pointing near the event
horizon.  At late times, after the merger, $f$ will be negative at C,
because C is inside the event horizon.  Throughout the whole
simulation, $\partial_i f = 0$ at C by symmetry, and therefore,
(\ref{e:VolumeMethod}) implies that $\partial_t f= 0$ there, so that
$f$ at C remains fixed at a finite negative value.  At merger,
however, the $f=0$ contour passes through $C$.  Therefore, $f$ must be
singular\footnote{Even with re-initializations of $f$, as performed
  in~\cite{Diener:2003}, the same argument applies to that time
  interval between re-initializations during which the topology of the
  EH changes.}.  Any method for solving the level-set equations that
assumes a smooth and regular solution (including finite-difference
methods that do not explicitly treat the singularity) will therefore
produce results that differ from the exact solution at the singular
point.  In~\cite{Diener:2003}, one-sided finite-difference stencils
are carefully chosen so as to not differentiate across the
singularity.

Because of these two issues we have stopped development of a spectral
implementation of the level-set method.  These problems arise because
of properties of the function $f$, which is merely a {\em tool} to
represent the actual surface of interest, $f=0$.  This surface itself
is well-behaved and smooth, suggesting it will be possible to evolve
this surface directly.  Geodesic and surface methods do precisely
this, and so we focus on these two methods in the remainder of this
paper.

\subsection{Numerical Implementation}
\label{ss:SpectralSurfaceImplementation}

Compared to the implementation of the geodesic method, implementing
the surface method is somewhat more complex due to the presence of
derivatives along the surface in (\ref{e:SurfaceSpatialDerivatives}).
Apart from this, geodesic method and surface method share rather
uniform implementation details.  We shall first discuss those aspects
that only apply to the surface method, and then follow with aspects
applicable to both methods.

We represent the surface $r^i(t, u, v)$ with spectral methods
(e.g.~\cite{Boyd1989}).  These methods approximate a desired function
$U({\mathbf x}, t)$ as a truncated expansion in basis functions
$\phi_k$, for instance Chebyshev polynomials or spherical harmonics:
\begin{equation}\label{eq:SpectralExpansion}
U(\boldsymbol{x}, t) = \sum_{k=0}^{N-1} \tilde{U}_k(t) \phi_k(\boldsymbol{x}),
\end{equation}
where $N$ is the order of the expansion.  The fundamental advantage of
spectral methods lies in their fast convergence: For smooth problems
and a suitable choice of basis functions, the error of the
approximation (\ref{eq:SpectralExpansion}) decreases {\em
  exponentially} with the number of basis functions per
dimension~\cite{Boyd1989}.  Derivatives of the function $U$ are
computed via the (analytically known) derivatives of the basis
functions.  Each set of basis functions has an associated set of
collocation points ${\mathbf x}_i$; a matrix multiplication translates
between function values at the collocation points, $U(\mathbf{x}_i)$,
and spectral coefficients $\tilde U_k$.

For the surface method, we represent each Cartesian component of
$r^i(t, u, v)$ (cf. (\ref{eq:Surface-Parameterization})) as an
expansion in scalar spherical harmonics,
\begin{equation}
\label{e:SurfaceYlms}
r^i(t,u,v) = \sum_{\ell=0}^L \sum_{m=-\ell}^{m=+\ell} \tilde A_{\ell
  m}^i(t)\;Y_{\ell m}(u,v).
\end{equation}
This expansion assumes that at fixed $t$, the surface has topology
$S_2$.  Note that (\ref{e:SurfaceYlms}) allows the surface to
intersect itself, as necessary in a binary merger for $t<t_{\rm CEH}$
(cf. Figure~\ref{fig:MultipleCross-sections}). Self-intersection is
possible because the coordinates $u$ and $v$ are {\em not} assumed to
be standard spherical angular coordinates, i.e. relations like
$\cos(u)= z/\sqrt{x^2+y^2+z^2}$ will in general not hold.

For spherical harmonics $Y_{\ell m}(u,v)$, the collocation points form
a rectangular grid in $(u, v)$, with the $u$ values chosen so that
$\cos(u)$ are the roots of the Legendre polynomial of order $L+1$, and
with the $v$ values being uniformly distributed in the interval
$[0,2\pi[$.  There are in total
\begin{equation}\label{eq:Npoints}
N=2(L+1)^2
\end{equation}
collocation points.  The evolution equations
(\ref{e:SurfaceMethodEvolutionEquation}) require derivatives
$\partial_u r^i$ and $\partial_v r^i$, which are computed by
transformation to spectral coefficients, application of recurrence
relations and inverse transform (using the SpherePack
library~\cite{spherepack-home-page}).  These derivatives are then
substituted into
(\ref{e:SurfaceMethodEvolutionEquation})--(\ref{e:Signoflambda}) to
compute $\partial_t r^i$, which is evolved at the collocation points.

We represent each Cartesian component $r^i$ as an expansion in scalar
spherical harmonics (see (\ref{e:SurfaceYlms})) in order to re-use the
infrastructure already developed for our spectral evolution code,
which represents tensors of arbitrary rank in this manner to simplify
our spectral expansions and to simplify communication of tensor
quantities across subdomains of different shapes (see, e.g.,
\cite{Holst2004,Scheel2006}).  An alternative approach would be to
represent $r^i$ in terms of {\em vector} spherical harmonics, i.e.,
\begin{equation}\label{eq:VectorSphericalHarmonic}
r^i(t,u,v) = \sum_{\ell=0}^L \sum_{m=-\ell}^{m=+\ell} \tilde
A_{\ell,m}(t)Y^i_{\ell m}(u,v).
\end{equation}
The downside of choosing a scalar spherical harmonic representation is
that the equation we impose on the highest order {\em vector}
spherical harmonics is incorrect, and this leads to an
instability. This difficulty with expanding vector quantities in a
scalar spherical harmonic basis is cured~\cite{Holst2004} by
performing the following ``filtering'' operation at each timestep:
first transform $r^i$ to a vector spherical harmonic basis, then
remove the $\ell = L$ and $\ell = L -1$ coefficients, and then
transform back.  The removal of both the highest and second highest
tensor harmonic modes is necessary, since transforming an n-th rank
tensor from a tensor spherical harmonics to scalar spherical harmonics
requires scalar harmonics of up to $L^{scalar} = L^{tensor} +n$.  We
filter two modes because we wish to correctly represent the spatial
derivatives of $r^i$ (see (\ref{e:SurfaceSpatialDerivatives})), which
are effectively rank 2.

The geodesic method simply evolves the ODEs
(\ref{e:GeodesicMethodEvolutionEquation2})-(\ref{e:GeodesicMethodEvolutionEquation1}).  While each geodesic is
evolved independently, we find it nevertheless convenient to represent
them as a two-dimensional grid, $q^i(t, u, v)$ where parameters $u$
and $v$ label each geodesic.  We use the same parameters $u$ and $v$
for geodesic and surface method, and for this paper, we choose to
locate the geodesics at the {\em same} $(u,v)$ values as the
collocation points of the surface method.  We note that this choice is
based on convenience to simplify comparison between the two methods;
geodesics can be placed at any location, and indeed, we plan as a
future upgrade of the geodesic method an adaptive placement of
geodesics to help resolve interesting features like caustics.

Let us now discuss aspects common to the implementation of the
geodesic and surface methods: At some late time $t=\tend$ long after
merger, we initialize the EH surface by choosing it to be the AH at
that time.  Our AH finder parameterizes the radius of the AH as a
function of standard azimuthal and longitudinal angles on $S_2$,
$r_{\rm AH}(\tend, \theta, \phi)$, i.e.
\begin{equation}
r^i_{\rm AH}(\tend,\theta,\phi)= r_{\rm AH}(\tend,\theta,\phi)
\left( \begin{array}{c} \sin \theta\cos \phi\\ \sin \theta\sin
  \phi\\ \cos \theta
 \end{array}
\right).
\end{equation}

When initializing the event horizon surface, we choose $(u, v)$ to
coincide with the standard spherical angular coordinates $(\theta,
\phi)$, i.e. we set
\begin{eqnarray}\label{eq:InitialData}
r^i(\tend, u, v) &=& r^i_{\rm AH}(\tend, u, v),\qquad\mbox{surface
  method},\\ q^i(\tend, u, v) &=& r^i_{\rm AH}(\tend, u,
v),\qquad\mbox{geodesic method}.
\end{eqnarray}
For the geodesic method we further set $p^i(\tend, u,v)=s^i_{\rm AH}$,
where $s^i_{\rm AH}$ is the unit normal to the apparent horizon, which
is computed similarly to
(\ref{e:SurfaceSpatialDerivatives})-(\ref{e:Signoflambda}).  Time
stepping is conducted using a 4th order Runge-Kutta algorithm.

Both methods require interpolation of certain quantities like the
spatial metric $\gamma_{ij}$ onto the grid points of the surface $r^i(t, u,
v)$.  For the spectral evolutions of the Caltech-Cornell
group~\cite{Pfeiffer-Brown-etal:2007,Boyle2007,Scheel2006} the
evolution data is represented as spectral expansions in space (for
each fixed time $t$) and spatial interpolation is performed by
evaluating the appropriate spectral expansions
(\ref{eq:SpectralExpansion}) at the desired spatial coordinates $r^i$.
Evolution data is available at discrete evolution times $t_n$ and
temporal interpolation is performed with 6-th order Lagrange
interpolation (i.e. utilizing 3 time slices on either side of the
required time).\footnote{The spectral spatial interpolation is
  computationally more expensive than temporal Lagrangian
  interpolation.  Whenever the domain decomposition for the Einstein
  evolution is identical for all timesteps involved in a temporal
  interpolation, the time interpolation is performed before the
  spatial interpolation.  In that case, only {\em one} spectral
  spatial interpolation is necessary (on the time-interpolated data),
  rather than six, speeding up the computation.}

Finally, we define an area element $\sqrt{h}$ on the surface as the
root of the determinant of the induced metric,
 \begin{eqnarray}\label{e:SurfaceAreaElement}
h = \frac{1}{\sin^2 u}\det \left( \begin{array}{cc} \gamma_{ij} \partial_u r^i
  \partial_u r^j & \gamma_{ij} \partial_u r^i \partial_v r^j
  \\ \gamma_{ij} \partial_v r^i \partial_u r^j & \gamma_{ij}
  \partial_v r^i \partial_v r^j \end{array} \right).
\end{eqnarray}
The area of the evolved surface is then given by
\begin{equation}
A(t) = \int dA = \int \sqrt{h(t,u,v)}\;\sin u\;
du\,dv. \label{e:SurfaceAreaIntegral}
\end{equation}
Explicitly pulling out the factor $\sin u$ in
(\ref{e:SurfaceAreaElement}) and (\ref{e:SurfaceAreaIntegral}) ensures
that $\sqrt{h}$ is a constant for a coordinate sphere in Euclidean
space; this will simplify Figure~\ref{fig:HeadonAreaElement} below.
Since all the geodesics (or surface grid points) are on a
Legendre-Gauss grid, we compute the derivatives in
(\ref{e:SurfaceAreaElement}) spectrally, and we evaluate
(\ref{e:SurfaceAreaIntegral}) by Legendre-Gauss quadrature.  For
binary black hole mergers before merger, we sometimes evaluate $h$
based on finite-difference derivatives $\partial_ur^i$ and $\partial_v
r^i$.  This is discussed in detail in
Section~\ref{ss:TreatmentOfMerger}.

\section{Application to Kerr spacetime}
\label{s:KerrApplication}

Initial tests of the event horizon finder are conducted using the Kerr
spacetime in Kerr-Schild coordinates (See \S33.6 of~\cite{MTW}):
\begin{equation}\label{eq:Kerr-Schild}
g_{\mu\nu} \equiv \eta_{\mu\nu} +2 H l_\mu l_\nu.
\end{equation}
Here $H$ is a scalar function of the coordinates, $\eta_{\mu\nu}$ is
the Minkowski metric, and $l^\mu$ is a null vector.  In Cartesian
coordinates $(t,x,y,z)$, the functions $H$ and $l^\mu$ for a black
hole of mass $M$ and dimensionless spin parameter $a/M$ in the $z$
direction are
\numparts
\begin{eqnarray}
H &=& {M r_{\rm BL}^3 \over r_{\rm BL}^4 + a^2 z^2},\\ l_\mu &=&
\left(1,{xr_{\rm BL}+ay\over r_{\rm BL}^2+a^2},{yr_{\rm BL}-ax\over
  r_{\rm BL}^2+a^2},{z\over r_{\rm BL}}\right),
\end{eqnarray}
\endnumparts
where $r_{\rm BL}(x,y,z)$ is the Boyer-Lindquist radial coordinate,
defined by
\begin{equation}
\label{eq:KerrRsquaredDef}
\fl \qquad r_{\rm BL}^2 = {1\over 2}\left(x^2 + y^2 + z^2 - a^2\right) + \left({1\over
  4}\left(x^2 + y^2 + z^2 - a^2\right)^2 + a^2 z^2\right)^{1/2}.
\end{equation}
If we define the Kerr-Schild spherical coordinates in the
straightforward way ($r = \sqrt{x^2+y^2+z^2}$, $\cos(\theta) = z/r$,
etc), we find that the event horizon of the Kerr black hole in these
coordinates is given by
\begin{equation}
r_{\rm Kerr}(\theta,\phi) = \sqrt{\frac{r_+^4 + r_+^2 a^2}{r_+^2 + a^2
    \cos^2\theta}},
\end{equation}
where $r_+ \equiv M+\sqrt{M^2-a^2}$.  The surface area of the event
horizon is given by
\begin{equation}
A_{\rm Kerr} = 8\pi M(M+\sqrt{M^2-a^2}).
\end{equation}

For our tests on Kerr spacetime we choose the same initial surface for
both the surface and geodesic methods: a coordinate sphere of radius
$r=2.5M$, which does not coincide with the horizon.  The evolution
begins at $\tend=0$ and proceeds backward in time towards negative
$t$.  Because we choose to place geodesics coincident with the
collocation points of the surface method (see
Section~\ref{ss:SpectralSurfaceImplementation}), we can use the
highest angular index $L$ as a measure of resolution.  The total
number of geodesics or grid points is given by (\ref{eq:Npoints}).
The choice of spin in the $z$ direction is for convenience.  We have
repeated the numerical tests below for spins of several different
orientations, and we find no substantial difference in either
stability or accuracy.

In order to test our methods of finding an EH, we use two measures of
error.  The first measures the error in the coordinate location of the
event horizon. We define
\begin{equation}
\label{eq:DeltaRErrorMeasure}
\Delta r(u,v) = r(u,v) - r_{\rm Kerr}(\theta(u,v),\phi(u,v)).
\end{equation} 
where $r(u,v)$, $\theta(u,v)$, and $\phi(u,v)$ are the Kerr-Schild
radial and angular coordinates of the surface, which are found from
either the surface-method variables $r^i(u,v)=[x(u,v),y(u,v),z(u,v)]$
or the geodesic-method variables $q^i(u,v)=[x(u,v),y(u,v),z(u,v)]$ in
the usual way, e.g., $x(u,v)=r(u,v)\sin\theta(u,v)\cos\phi(u,v)$.
Specifically, we will use the root-mean-square of $\Delta r$ over all
grid points or geodesics, which we shall denote by $||\Delta r||$, as
a global measure of the error.

\begin{figure}
\centerline{\includegraphics[width=.9\textwidth]{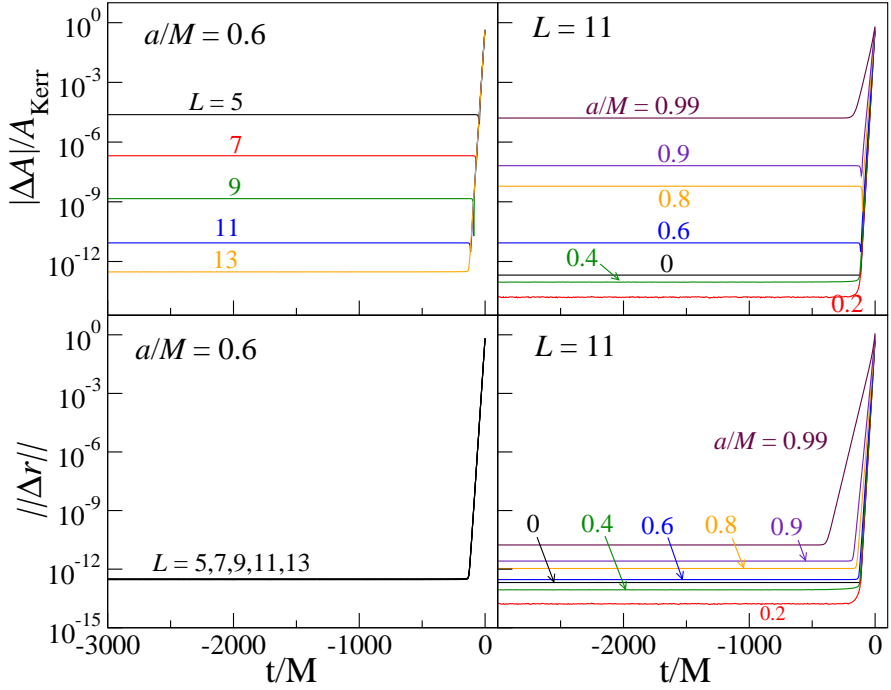}}
\caption{{\bf Geodesic method} applied to a Kerr black hole.  The {\bf
    top panels} show the area difference between the computed and
  exact solution, normalized by the area of the exact solution.  The
  {\bf bottom panels} show the difference between the computed and
  exact location of the EH, as measured by
  (\ref{eq:DeltaRErrorMeasure}).  These data are shown for two series
  of runs: In the {\bf left panels} we keep the dimensionless spin of
  the black hole fixed at $a/M=0.6$ and vary the resolution $L$ of
  the EH finder.  In the {\bf right panels} we vary the spin parameter
  $a/M$ at fixed resolution.  In all cases, the EH finder starts at
  $t=0$ and the geodesics are evolved backward in time.
\label{fig:Geodesic4GraphKerr}}
\end{figure}

Our second error measure is the deviation of the area of our surface
  from the Kerr value,
\begin{equation}
\label{eq:DeltaAErrorMeasure}
\Delta A = A(t) - A_{\rm Kerr},
\end{equation}
where $A(t)$ is determined by Equation~(\ref{e:SurfaceAreaIntegral}).

Figure~\ref{fig:Geodesic4GraphKerr} shows errors in the AH surface as
computed using the geodesic method for a Kerr black hole.  The error
measure $||\Delta r||$, (\ref{eq:DeltaRErrorMeasure}),
does not change with $L$ because the evolution of each geodesic is
independent of the total number of geodesics. The error measure
$|\Delta A|$, (\ref{eq:DeltaAErrorMeasure}), does depend on $L$, but
only because the computation of the surface area depends on all
geodesics.  It is clear from Figure~\ref{fig:Geodesic4GraphKerr} that
the geodesic method can stably model Kerr black holes of any spin.

\begin{figure}
\centerline{
  \includegraphics[width=0.6\textwidth]{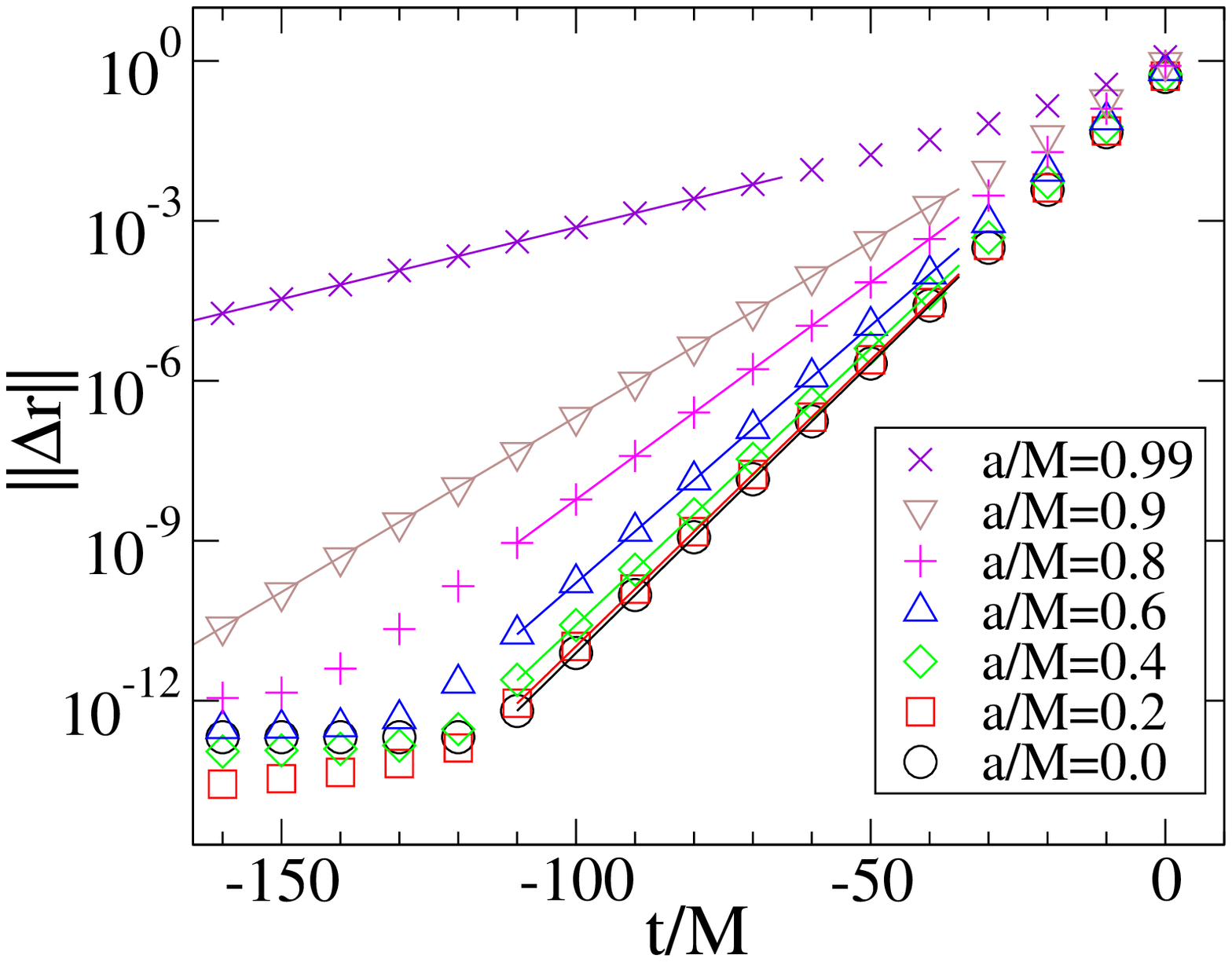} }
\caption{
\label{fig:KerrSurfaceGravity} Approach of the
tracked null surface onto the event horizon of Kerr black holes with
various spins.  The symbols show the numerical data (the same data as
in the lower right panel of Figure~\ref{fig:Geodesic4GraphKerr}), and
the solid lines are representative least-squares fits.
Table~\ref{tab:SurfaceGravity} compares the numerically computed
e-folding time to the surface gravity of the black hole.}
\end{figure}

At $\tend=0$, we start the EH finder with an initial surface that does
{\em not} coincide with the EH of Kerr.  Therefore,
Figure~\ref{fig:Geodesic4GraphKerr} shows initial transients as the
surface being followed by the EH finder approaches the EH of Kerr.
Figure~\ref{fig:KerrSurfaceGravity} shows an enlargement of this
phase.  We find that the tracked surface approaches the Kerr EH
exponentially when integrating backward in time,
\begin{equation}
||\Delta r|| \propto e^{\,t/\tau}.
\end{equation}
The time scale $\tau$ depends on the spin of the Kerr background.
It has been shown in a number of coordinate
  systems~\cite{Libson96,Caveny2003a,Diener:2003} that the e-folding
  time for a non-spinning black hole is $\tau = 4M$.  This is not true
  in all coordinate systems: for example, in Schwarzschild coordinates
  $\tau = 2M$.  In~\ref{App:SurfaceGravity}, we generalize this
  result to show that null geodesics, perturbed from the Kerr EH,
diverge from the EH exponentially with an e-folding time equal to
$1/\gH$, where
\begin{equation}
\gH=\frac{\sqrt{M^2-a^2}}{2M\left(M+\sqrt{M^2-a^2}\right)}
\end{equation}
is the surface gravity of the horizon in Kerr-Schild coordinates.  In
Table~\ref{tab:SurfaceGravity}, we compare the numerically computed
e-folding time $\tau$ (obtained by least-squares fits) to $\gH$, and
find excellent agreement.

\begin{table}
\caption{\label{tab:SurfaceGravity} Exponential approach of the null
  surface to the correct event horizon location.  $M\gH$ represents
  the (dimensionless) surface-gravity of a Kerr black hole with spin
  $a/M$.  $M/\tau$ is the numerical rate of approach as determined by
  fits to the data shown in Figure~\ref{fig:KerrSurfaceGravity}.  }

\centerline{\begin{tabular}{|c|ccr|}
\hline
$a/M$  &  $M\gH$  &  $M/\tau$  &  $M\gH-M/\tau$ \\\hline
0.0  & $1/4=0.25$    & 0.249998 & $2\cdot 10^{-6}$ \\
0.2  & 0.247449 & 0.247440  & $9\cdot 10^{-6}$ \\ 
0.4  & 0.239110 & 0.239093 & $1.7\cdot 10^{-5}$ \\
0.6  & 0.222222 & 0.222212 & $1.0\cdot 10^{-5}$ \\
0.8  & $3/16=0.1875$   & 0.187500   & $<10^{-6}       $ \\
0.9  & 0.151784 & 0.151784 & $<10^{-6}       $ \\
0.99 & 0.061814 & 0.061814 & $<10^{-6}       $ \\ \hline
\end{tabular}}
\end{table}

\begin{figure}
\includegraphics[width=0.48\textwidth]{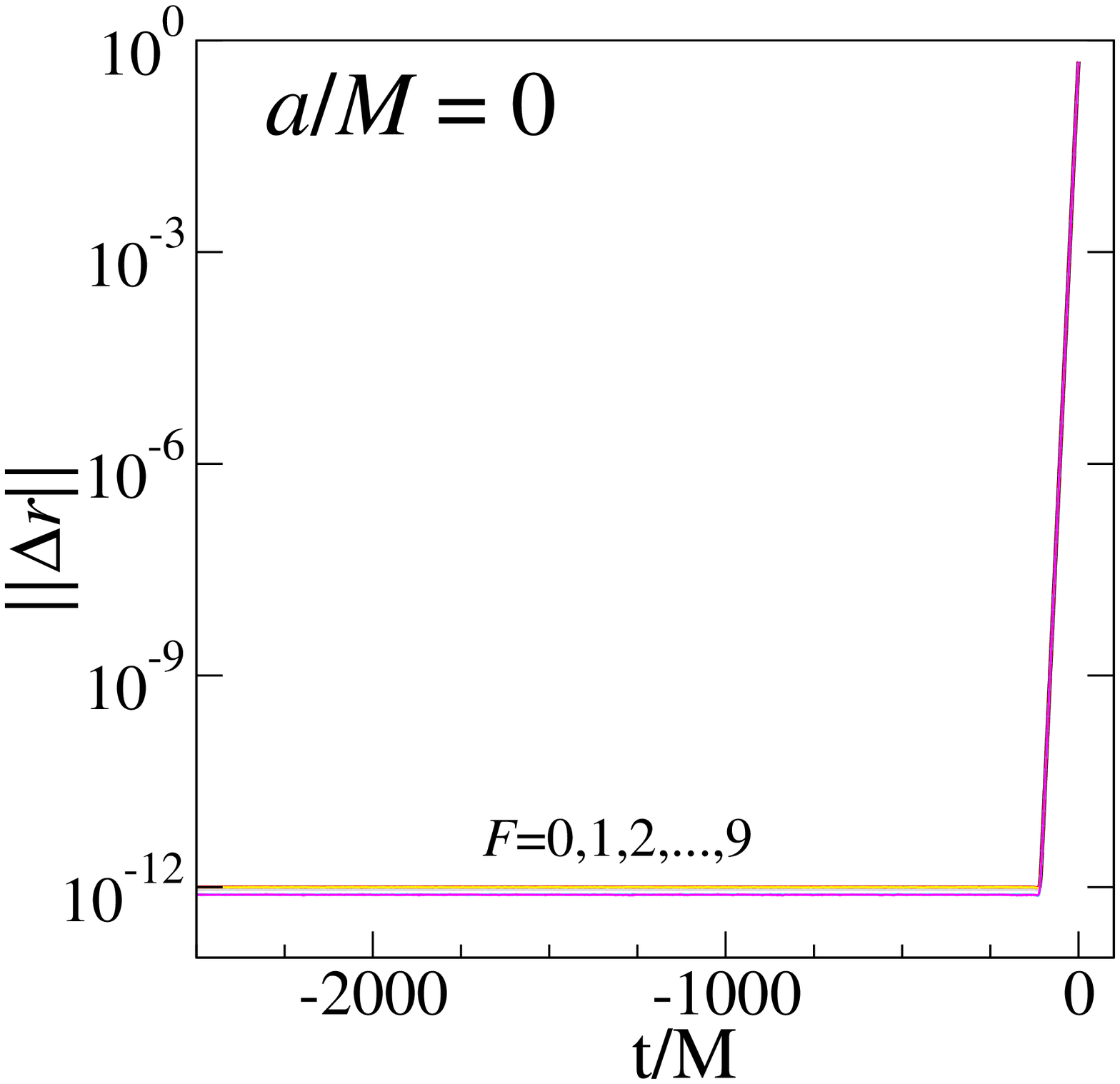}
\hspace{0.02in}
\includegraphics[width=0.48\textwidth]{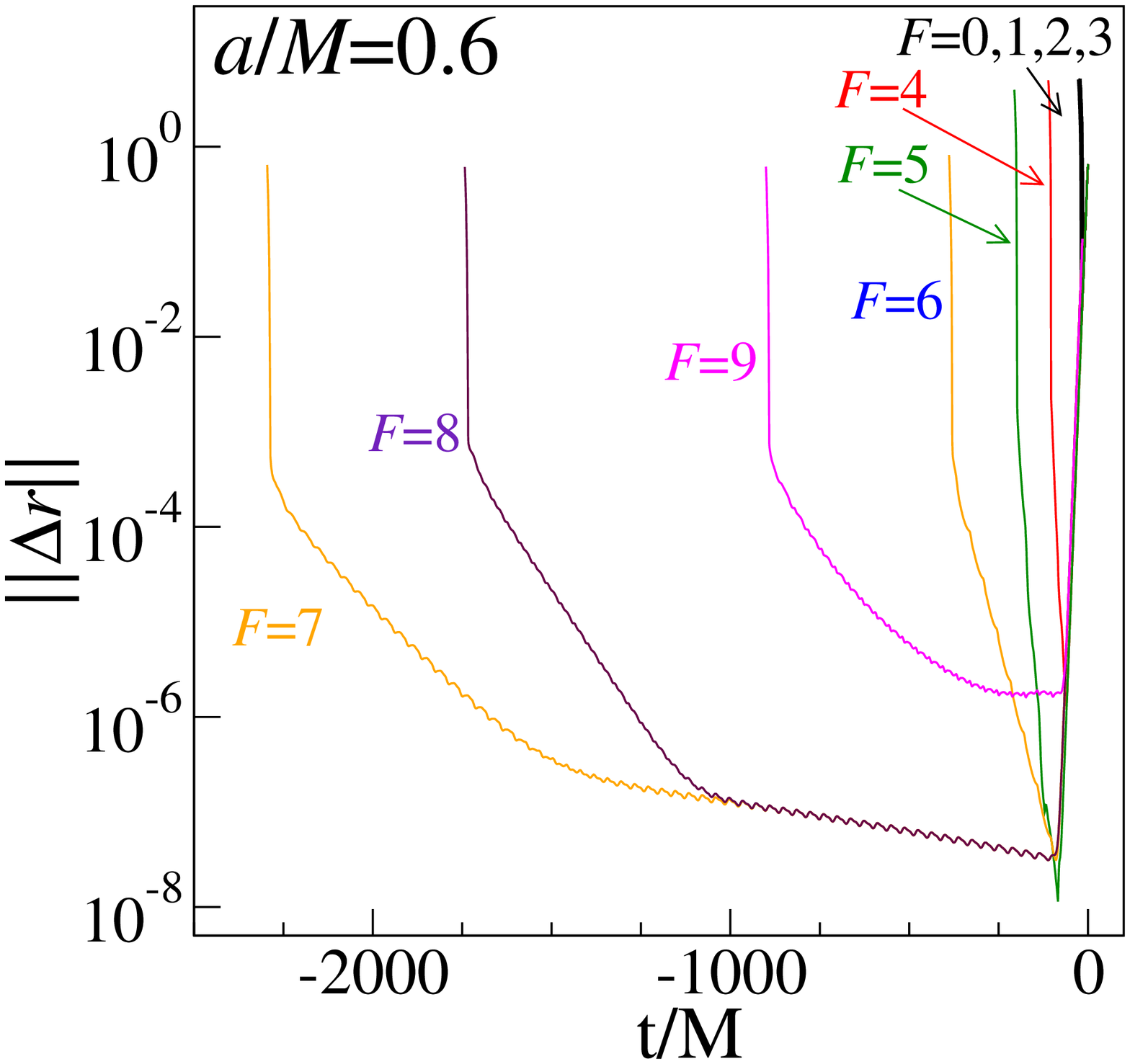}
\caption{Effect of filtering using (\ref{eq:RadialFiltering}) and
  (\ref{eq:RadialFiltering2}) for the {\bf surface method}.  Shown are
  evolutions with the same angular resolution $L=18$, but for
  different numbers $F$ of truncated modes in
  (\ref{eq:RadialFiltering}).  {\bf Left panel:} For a Schwarzschild
  black hole, the surface method is stable with or without this
  additional filtering.  {\bf Right panel:} For a Kerr black hole with
  $a/M=0.6$, $F=7$ performs best.  The EH finder starts at $t=0$ and
  the surface is evolved backward in time.
\label{fig:SurfaceFilteringKerr}}
\end{figure}

We now turn our attention to the surface method.  For a Schwarzschild
black hole, the surface method with the standard tensor spherical
harmonic filtering is stable, as shown by the ``F=0'' line in the left
panel of Figure~\ref{fig:SurfaceFilteringKerr}.
However, the method is unstable for spinning black holes and fails
within about $10M$ for spin $a/M=0.6$ (see the ``F=0'' line in the
right panel of Figure~\ref{fig:SurfaceFilteringKerr}).

Therefore, we perform additional filtering for spinning black holes.
After each timestep, we compute
\begin{equation}
R(u,v)= \sqrt{\delta_{ij} r^i(u,v) r^j(u,v)},
\end{equation}
expand $R(u,v)$ in scalar spherical harmonics,
\begin{equation}
R(u,v)=\sum_{\ell=0}^L\sum_{m=-\ell}^\ell \tilde R_{\ell m}Y_{\ell m}(u,v),
\end{equation}
and truncate the highest $F$ modes of this expansion:
\begin{equation}\label{eq:RadialFiltering}
\tilde R_{\ell m} \to 0,\quad\mbox{for $\ell > L-F$}.
\end{equation}
From these filtered coefficients, we reconstruct the filtered
radius-function $R_F(u,v)$ and replace
\begin{equation}\label{eq:RadialFiltering2}
r^i \rightarrow \frac{R_F}{R}r^i.
\end{equation}

The right panel shows that with appropriate choice of $F$, the horizon
of a Kerr black hole with spin $a/M=0.6$ can be followed for thousands
of $M$.  Unfortunately, we do not understand the effect of $F$ on
stability, and therefore a parameter search through possible values
for $F$ is required.

\begin{figure}
\centerline{\includegraphics[width=0.9\textwidth]{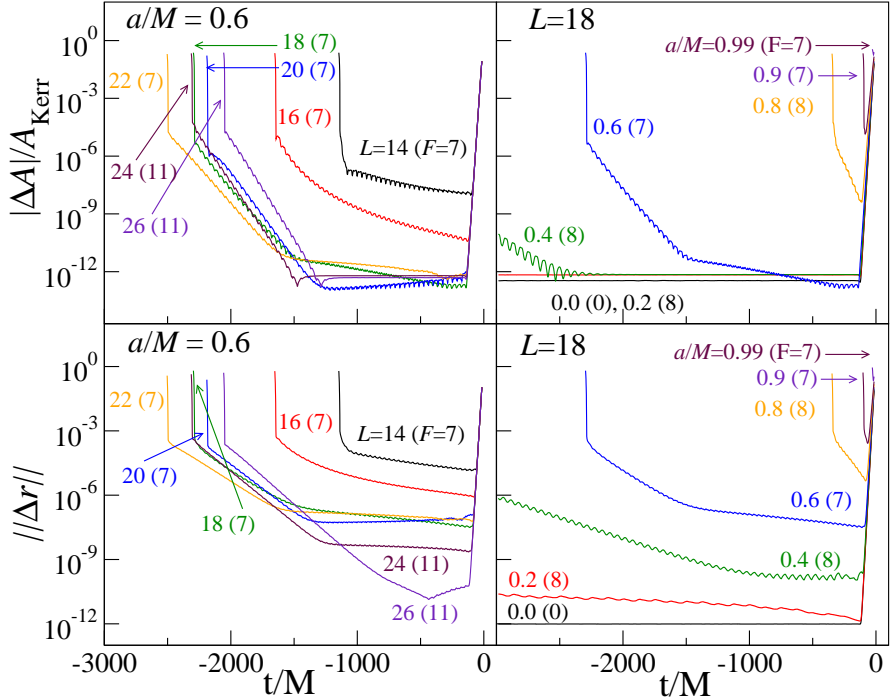}}
\caption{{\bf Surface Method}, applied to a Kerr black hole.  The {\bf
    top panels} show the normalized area difference between the
  computed and exact solution.  The {\bf bottom panels} show the
  difference between the computed and exact location of the EH, as
  measured by (\ref{eq:DeltaRErrorMeasure}).  These data are shown for
  two series of runs: In the {\bf left panels} we keep the
  dimensionless spin of the black hole fixed at $a/M=0.6$ and vary the
  resolution $L$ of the EH finder.  In the {\bf right panels} we vary
  the spin $a$ at fixed resolution.  The value $F$ denotes the number
  of truncated modes during filtering according to
  (\ref{eq:RadialFiltering}).  For each case, we show the value of $F$
  that provides the most accurate evolution.  Also, in all cases, the
  EH finder starts at $\tend=0$ and the surface is evolved backward in
  time. Compare to Figure~\ref{fig:Geodesic4GraphKerr}.
\label{fig:Surface4GraphKerr}}
\end{figure}

With this additional filtering in place, we now examine the
convergence and accuracy of the surface method.
Figure~\ref{fig:Surface4GraphKerr} shows the convergence behaviour of
the surface method.  From the top plots, we can see that for a black
hole of moderate spin ($a/M=0.6$), the surface method is accurate and
convergent, although long-term stability issues remain.  Also, the
surface area computed by the surface method appears to be more
accurate than the location of the surface, cf. upper vs. lower panels
of Figure~\ref{fig:Surface4GraphKerr}.  This arises, because for a
small change $\delta \tilde A^i_{lm}$ in an expansion coefficient
$\tilde A^i_{lm}$ in (\ref{e:SurfaceYlms}) with $\ell\neq 0$, the
change in $||\Delta r||$ is linear in $\delta \tilde A_{lm}^i$,
whereas the change in area is quadratic.  The high accuracy of $A_{\rm
  EH}$ is a welcome feature, since the EH area is one of the most
important results of an EH finder.  Unfortunately, the surface method
is not capable of tracking the horizon for spins $a/M\gtrsim 0.8$ for
a useful length of time.

While the geodesic method appears superior in these Kerr tests, there
are two main benefits to implementing the surface method.  Firstly, it
is computationally more efficient.  The bulk of processing time is
spent on interpolating the metric data from the simulation, and the
surface method requires the metric only (10 components) whereas the
geodesic method requires the metric, as well as its spatial and time
derivatives (50 components).  Secondly, the surface method can be used
to check the errors in the geodesic method in circumstances where the
surface method performs well, i.e. lower spins.

For these tests, the initial set of geodesics (or surface) is chosen
to be a sphere of radius 2.5M.  In this case it requires a time
$\gtrsim 100M$ for either method to converge onto the actual event
horizon.  This shows that for cases in which the actual EH is unknown,
it is important to have a near-stationary situation at the end of the
simulation, so that the initial guess (generally taken to be the AH)
has time to converge onto the true EH.  The length of this interval
will depend on the desired accuracy, the quality of the initial guess
and the spin of the black hole.  For example, during a time $\Delta
t=10/\gH$ (i.e. $40M$ for $a/M=0$, but $160M$ for $a/M=0.99$) the
tracked surface will have approached the EH to a fraction $e^{-10}
\simeq 5\cdot 10^{-5}$ of the distance between the initial guess and
the EH.

\section{Head-on Binary Black Hole Merger}
\label{s:HeadonMergerApplication}

\subsection{Details of BBH evolution}
\label{ss:HeadonSimulation}

When looking for a straightforward dynamical spacetime where tracking
the event horizon is of interest, one of the standard scenarios is the
head on merger of two equal-mass non-spinning black holes
\cite{Libson95a,Libson96,matzner_etal95,MassoEtAl:1999,Caveny2003a}.
First, the SpEC code is utilized to evolve the solution of Einstein's
equations for the head-on merger.  Initially the holes are at rest, $r
\simeq 4.5M$ apart, where $M=M_A+M_B$
is the total mass at $t=0$ (because the black holes are non-spinning,
we take the irreducible mass as the black hole mass, $M_{A/B}=M_{{\rm
    irr}\,A/B}=\sqrt{A_{{\rm AH}\,A/B}/(16\pi)}$).  Initial data is
constructed by solving the conformal thin sandwich
equations~\cite{York1999,Pfeiffer2003b} with the same setup as
in~\cite{Scheel2006}, but setting the orbital frequency $\Omega_0=0$.
This data is then evolved with the SpEC code using the dual coordinate
frame technique described in~\cite{Scheel2006} and with a domain
decomposition with two excision spheres.  A common apparent horizon
appears at $t=t_{\rm CAH}=17.83M$.  Shortly thereafter, at $t=t_{\rm
  regrid}=18.96M$, the original domain decomposition with two excision
boundaries is replaced by a set of concentric spherical shells with
one larger excision boundary.  The new excision boundary lies somewhat
inside the common apparent horizon, but outside the original excision
boundaries.  The region very close to the original excision
boundaries, and between them, is dropped, and is no longer evolved.
Data is interpolated from the highest resolution merger run onto three
resolutions of this new domain decomposition.  The simulation is
continued up to $t=95M$ and the final mass of the merged black hole is
$\Mfinal=0.9493 M$.

\begin{figure}
\includegraphics[width=0.48\textwidth]{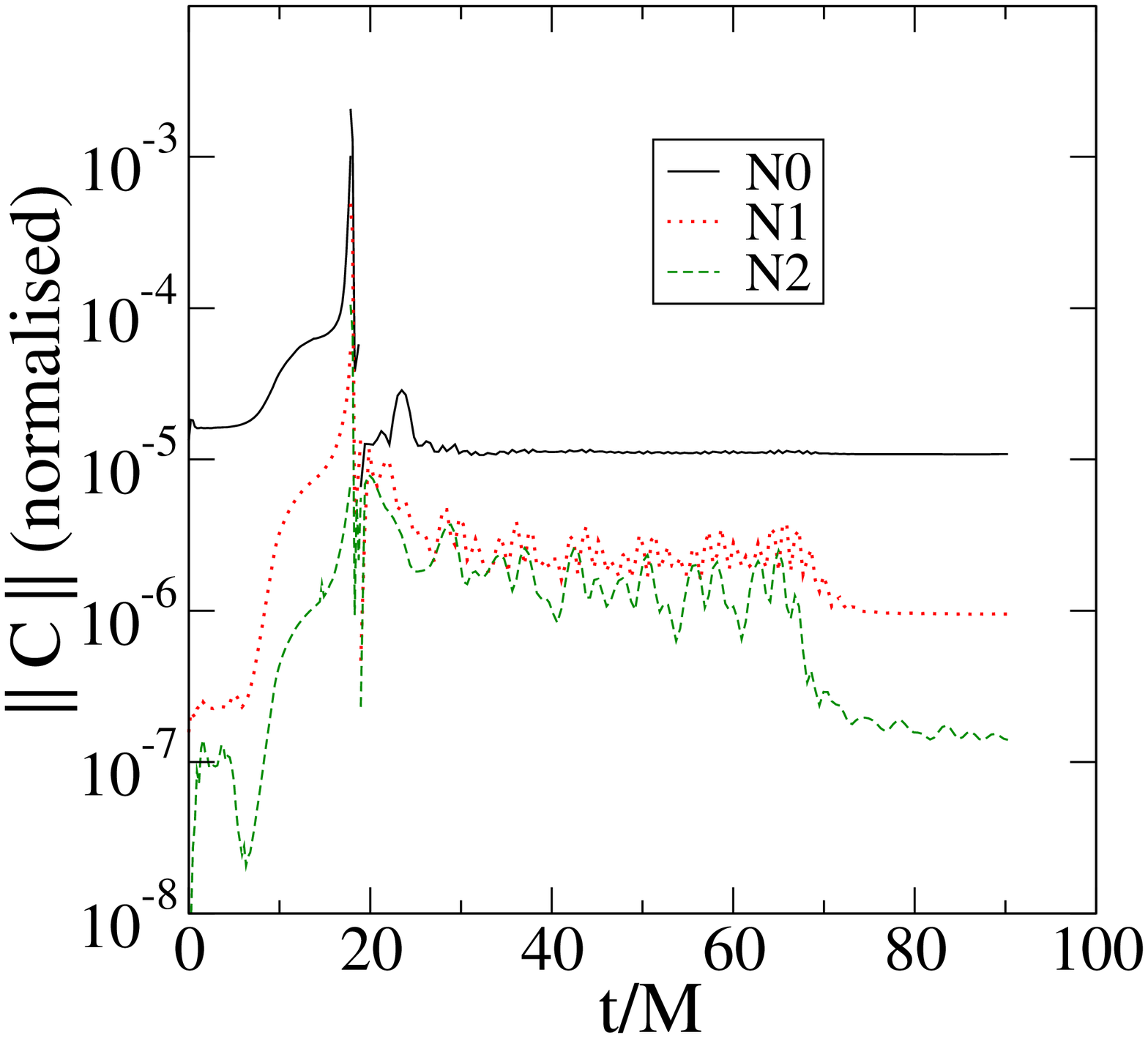}
\hspace{0.02in}
\includegraphics[width=0.48\textwidth]{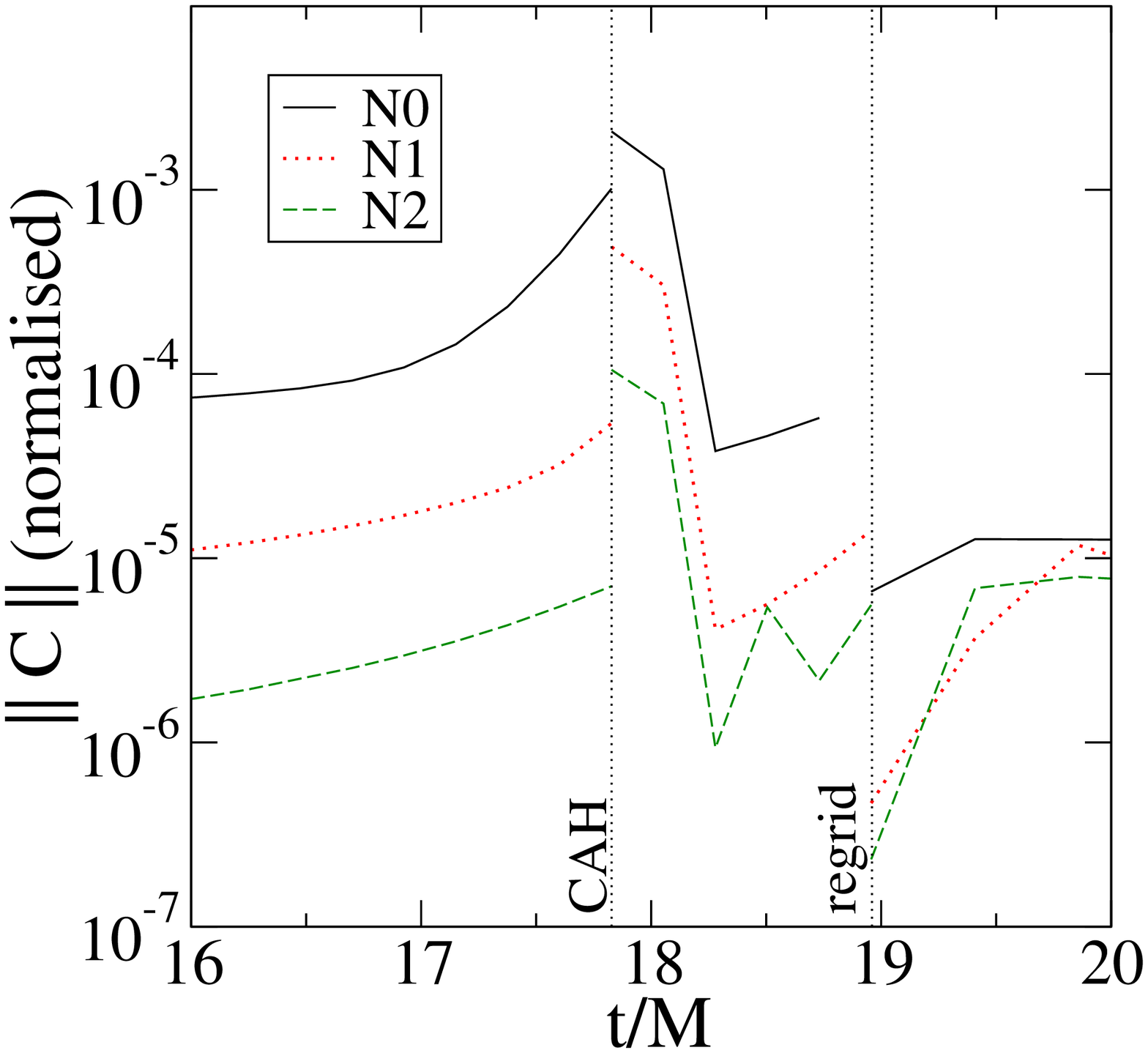}
\caption{Evolution of a head-on BBH merger: {\bf normalized constraint
    violations}.  The left panel shows the complete evolution.  The
  right panel enlarges the time around merger, with formation of a
  common apparent horizon and time of regridding indicated by 'CAH'
  and 'regrid', respectively.  The discontinuity at $t_{\rm CAH}$
  arises because the constraints are computed only outside the common
  AH for $t>t_{\rm CAH}$.  At $t_{\rm regrid}$, the constraints jump
  because of the different numerical truncation error of the ringdown
  domain decomposition.
\label{fig:AxiSymmetricConstraints}}
\end{figure}

The simulation is performed at three progressively higher resolutions,
named 'N0' through 'N2'.  The SpEC code does not strictly enforce the
Hamiltonian or momentum constraints, nor the artificial constraints
that arise from the first-order reduction of the Generalized Harmonic
formulation of Einstein's equations~\cite{Lindblom2006}.  As such, it
is important to monitor the values of these constraints during the
simulation, as shown in Fig.~\ref{fig:AxiSymmetricConstraints}.  We
normalize the constraints by an appropriate norm of the derivatives of
the evolved variables (see (71) of~\cite{Lindblom2006} for the precise
definition) and integrate constraint violations and normalization only
outside the two individual apparent horizons or the common apparent
horizon for this run.


\subsection{EH finder behaviour}
\label{ss:EHFinder_behaviour}

Since the EH finder follows the EH backward in time, we begin our
discussion with the ringdown phase of the head-on merger.  Initial
data for both the geodesic and surface methods is taken from the
apparent horizon at $t = 81.24M$, about $60M$ after appearance of a
common AH.

We run both the geodesic and surface methods for angular resolutions
$L=7,15,23,\ldots,47$ and compute the area $A(t)$ of the tracked
surface for these runs.  We do not employ filtering as
per~(\ref{eq:RadialFiltering}) for the surface method.

\begin{figure}
\centerline{
\includegraphics[width=0.48\textwidth]{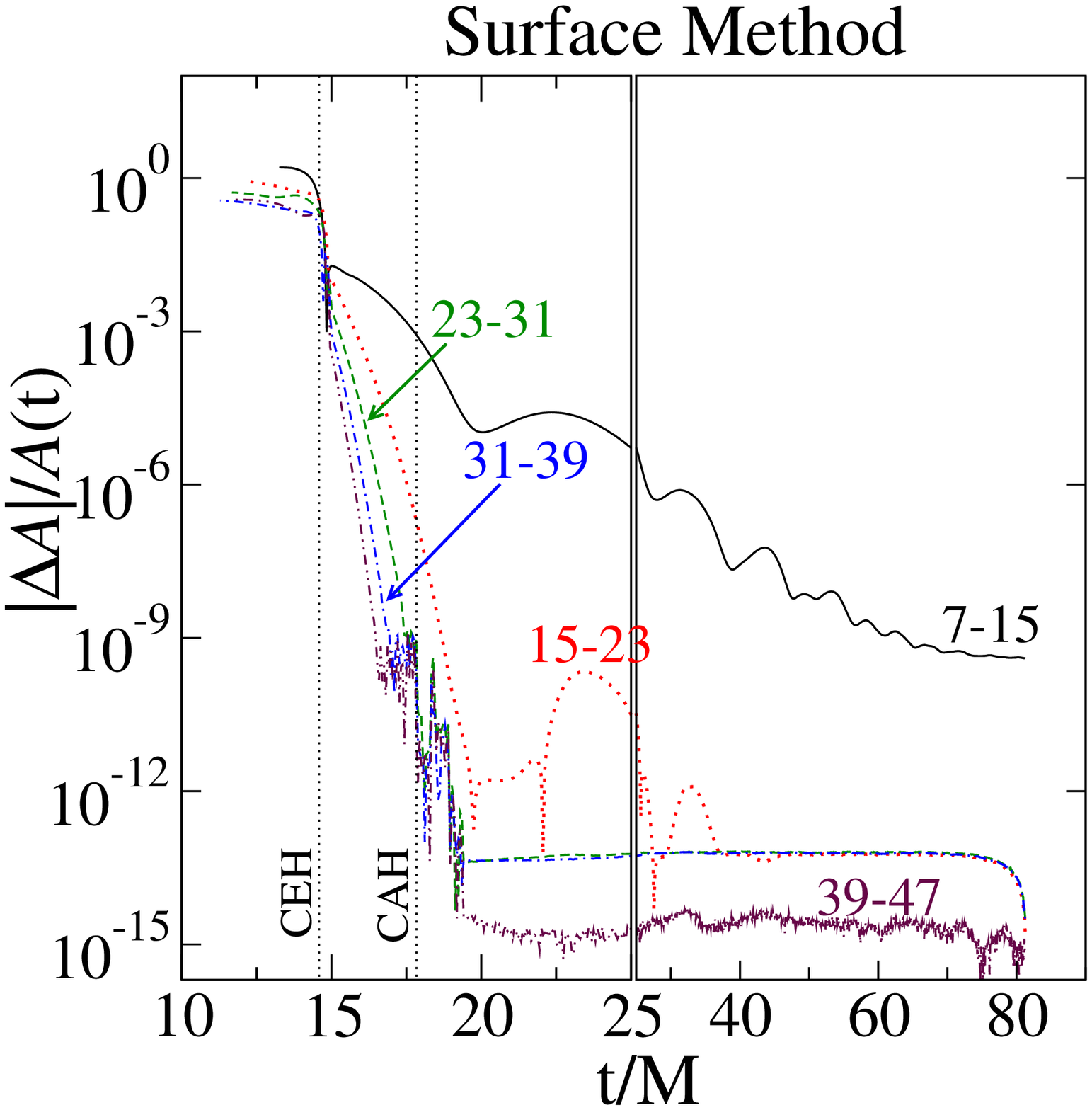}
\hspace{0.02in}
\includegraphics[width=0.48\textwidth]{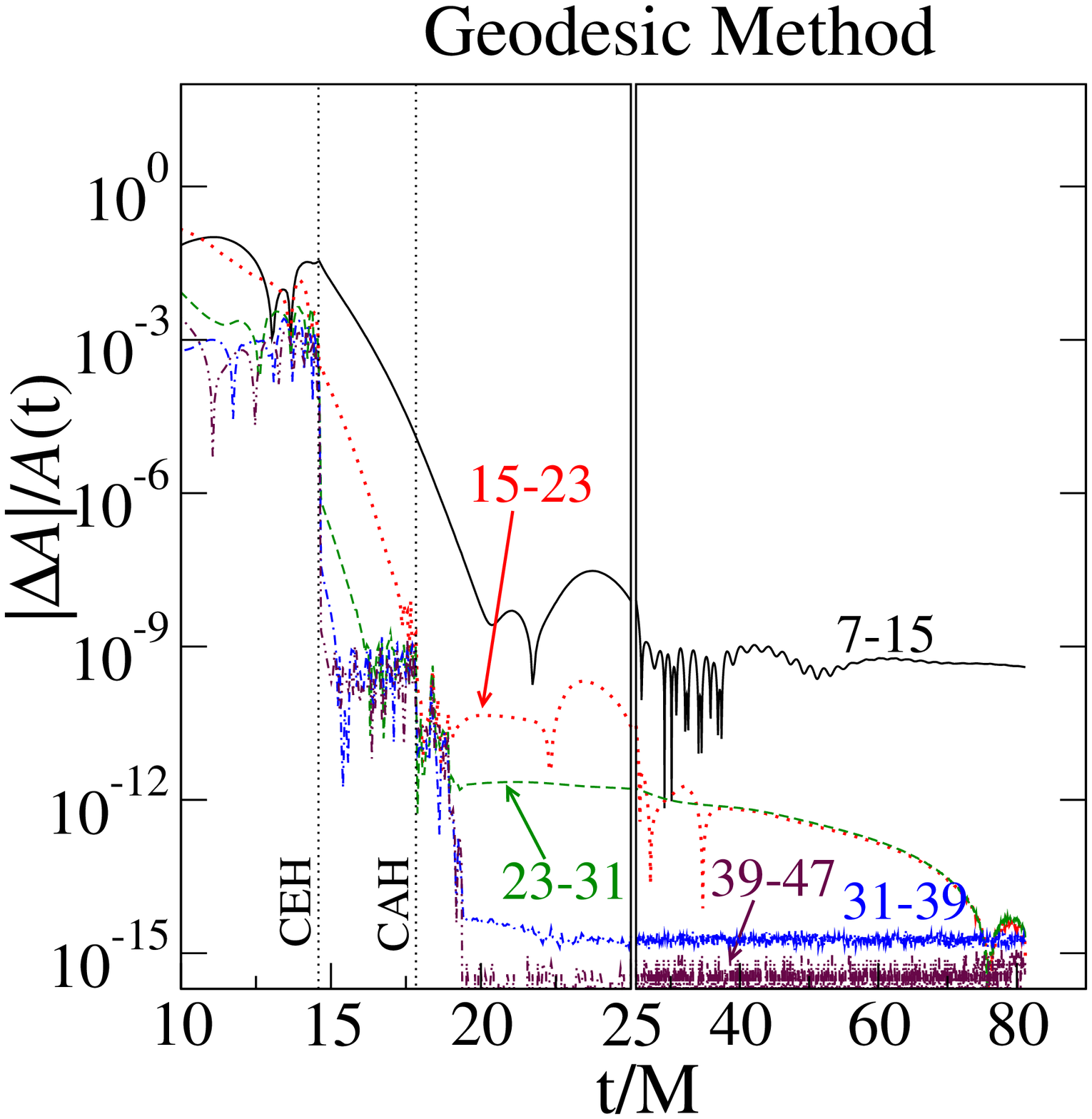}
}
\caption{Effect of changing the resolution of the EH finder when
  applied to the BBH evolution at fixed high resolution.  Shown are
  relative differences in the area $A(t)$ of the tracked surface.  The
  label ``7-15'' denotes the difference between simulations with
  $L_1=7$ and $L_2=15$, normalized by $A(t)$ of $L_2$.  Vertical lines
  on the graph denote the formation of common event and apparent
  horizons.  Note that the time scale of both plots change at
  $t/M=25$.
\label{fig:EHSEHGConvergence}}
\end{figure}

Figure~\ref{fig:EHSEHGConvergence} plots the relative differences
between $A(t)$ computed with different angular resolution. This plot
exhibits several noteworthy features, which we discuss in the next few
paragraphs:

During the ringdown phase, $t\gtrsim 20M$, both the surface and
geodesic methods perform admirably: Even at low resolution $L=7$, the
area is computed to better than $10^{-6}$ and this error drops rapidly
below $10^{-12}$ as $L$ is increased.  The rapid convergence with $L$
in the ringdown regime is not too surprising, because the angular
resolution of the merger simulation is $L_{\rm evolution} = 25$.
Therefore, angular modes $\ell>25$ of the EH finder carry only
information about the way in which the surface parameters $(u,v)$
deviate from the $(\theta,\phi)$ coordinates of the simulation.  As
can be seen from the excellent convergence for $t\gtrsim 20M$ in
Figure~\ref{fig:EHSEHGConvergence}, such deviations are not very
important.  We also note that the long-term instability exhibited by
the surface method during the Kerr test is not apparent.

Close to merger and before merger, $t\lesssim 20M$, the tracked
surface becomes very distorted and therefore requires much higher
angular resolution.  This is apparent in the comparatively larger
errors in $A(t)$ for $t_{\rm CEH}<t\lesssim 20M$.  In this time
interval, the errors in the surface method grow more rapidly than
those of the geodesic method.  We attribute this to a degradation of
the convergence rate of the spectral expansion (\ref{e:SurfaceYlms}).
The surface method relies on the spectral expansion in an essential
way to compute the derivatives that enter into
(\ref{e:SurfaceSpatialDerivatives}).  In contrast, evolution of
geodesics is independent of the spectral expansion and the spectral
series is used only to compute the surface area via
(\ref{e:SurfaceAreaIntegral}).

At the point of merger, when the surface being tracked by the EH
finders intersects itself for the first time, the error in the
area-computation suddenly increases drastically in either method.  The
reasons for this are quite different for the two methods: The geodesic
method evolves individual geodesics perfectly fine through $t_{\rm
  CEH}$.  The large errors in Figure~\ref{fig:EHSEHGConvergence} arise
because of the use of spectral integration to compute the surface
area: At a caustic, the surface-area element $\sqrt{h}$,
(\ref{e:SurfaceAreaElement}), tends to zero, resulting in a non-smooth
integrand in the area integral (\ref{e:SurfaceAreaIntegral}),
destroying exponential convergence of the spectral area integration.
Below, we will explain how we employ finite-difference integration
instead.  We shall address area calculation for $t<t_{\rm CEH}$ in
Section~\ref{ss:TreatmentOfMerger}, where we also discuss how to
compute the area of the EH excluding the future generators of the EH.

The surface method exhibits additional, more fundamental, problems at
$t_{\rm CEH}$, when the surface being tracked intersects itself in a
caustic with $\sqrt{h}\to 0$.  At such a point, the tangents to the
surface, $\partial_ur^i$ and $\partial_v r^i$ are either no longer
linearly independent, or one of them is zero,
cf. (\ref{e:SurfaceAreaElement}).  Therefore the surface normal $s^i$
in (\ref{e:SurfaceSpatialDerivatives}) is ill-defined.

\begin{figure}
\includegraphics[width=0.48\textwidth]{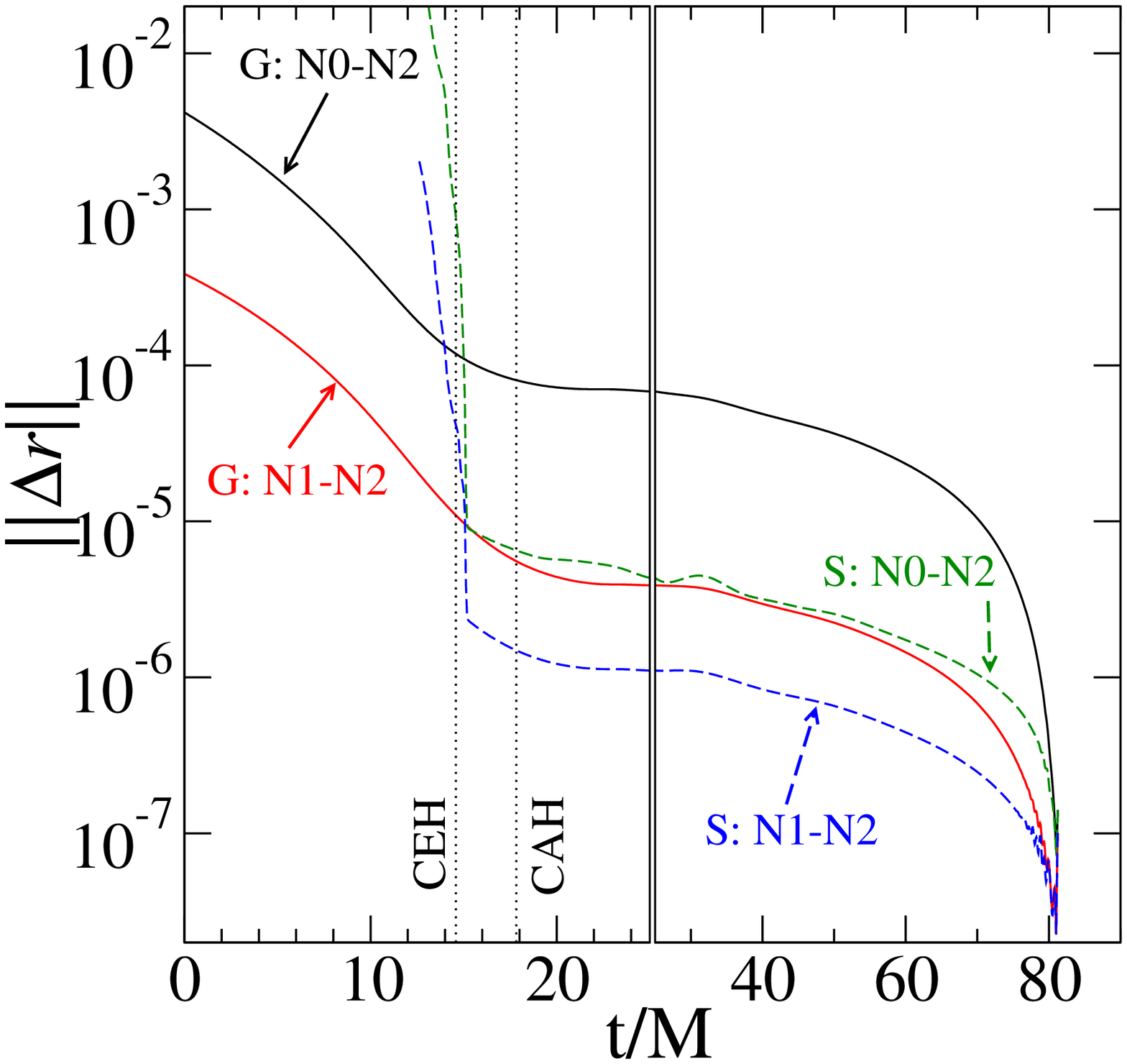}
\hspace{0.02in} \includegraphics[width=0.48\textwidth]{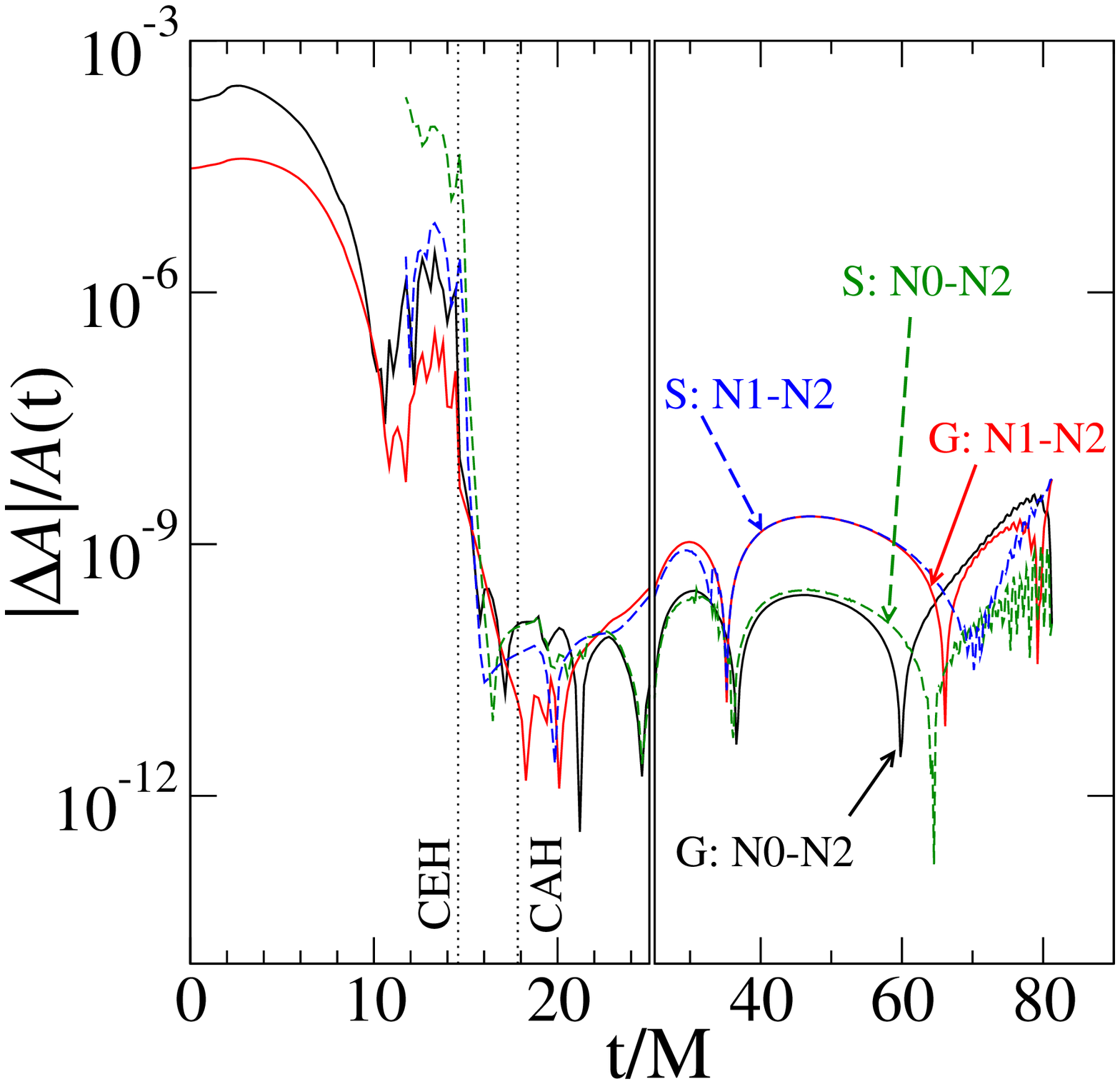}\\
\includegraphics[width=0.48\textwidth]{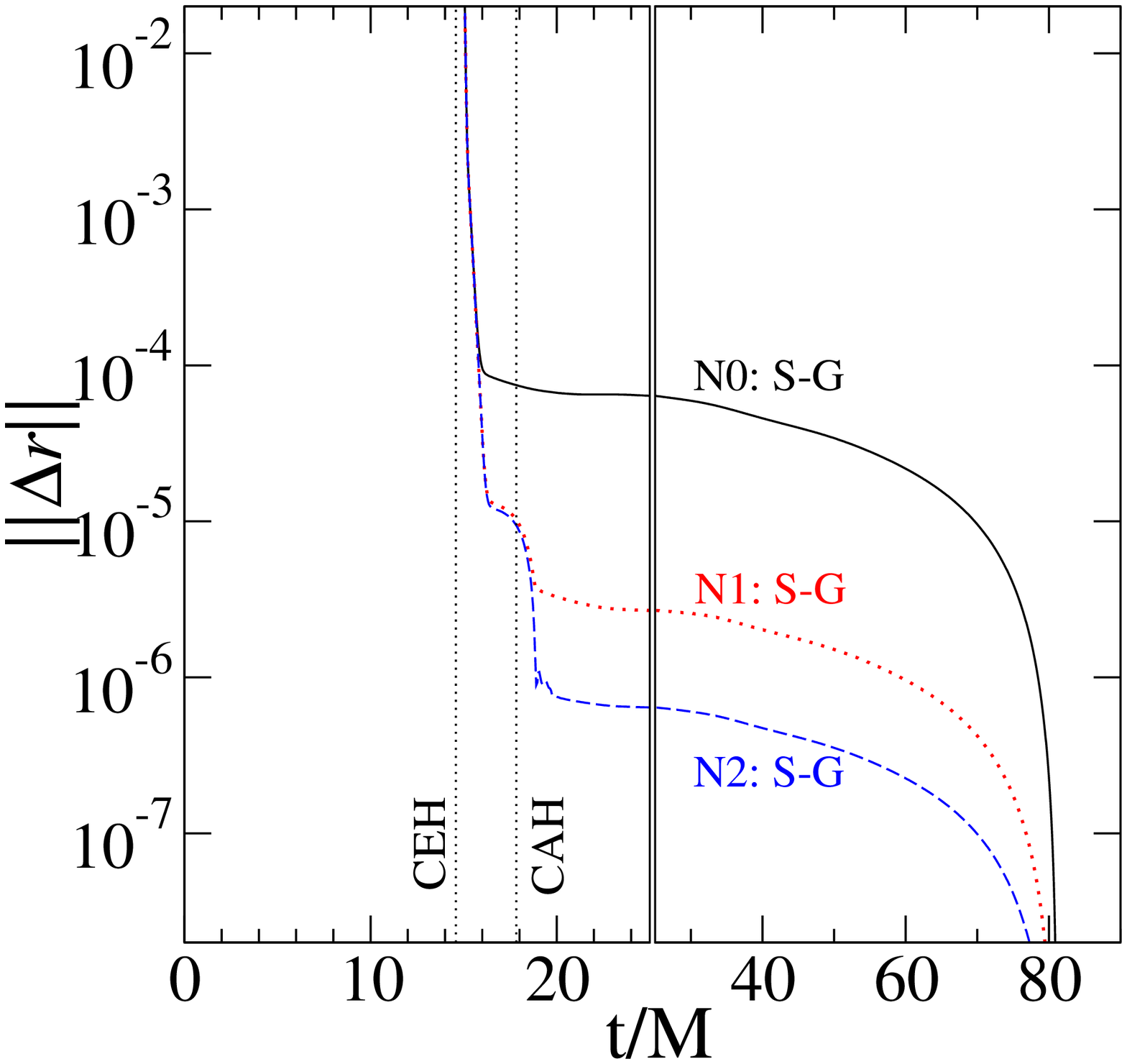}
\hspace{0.02in} \includegraphics[width=0.48\textwidth]{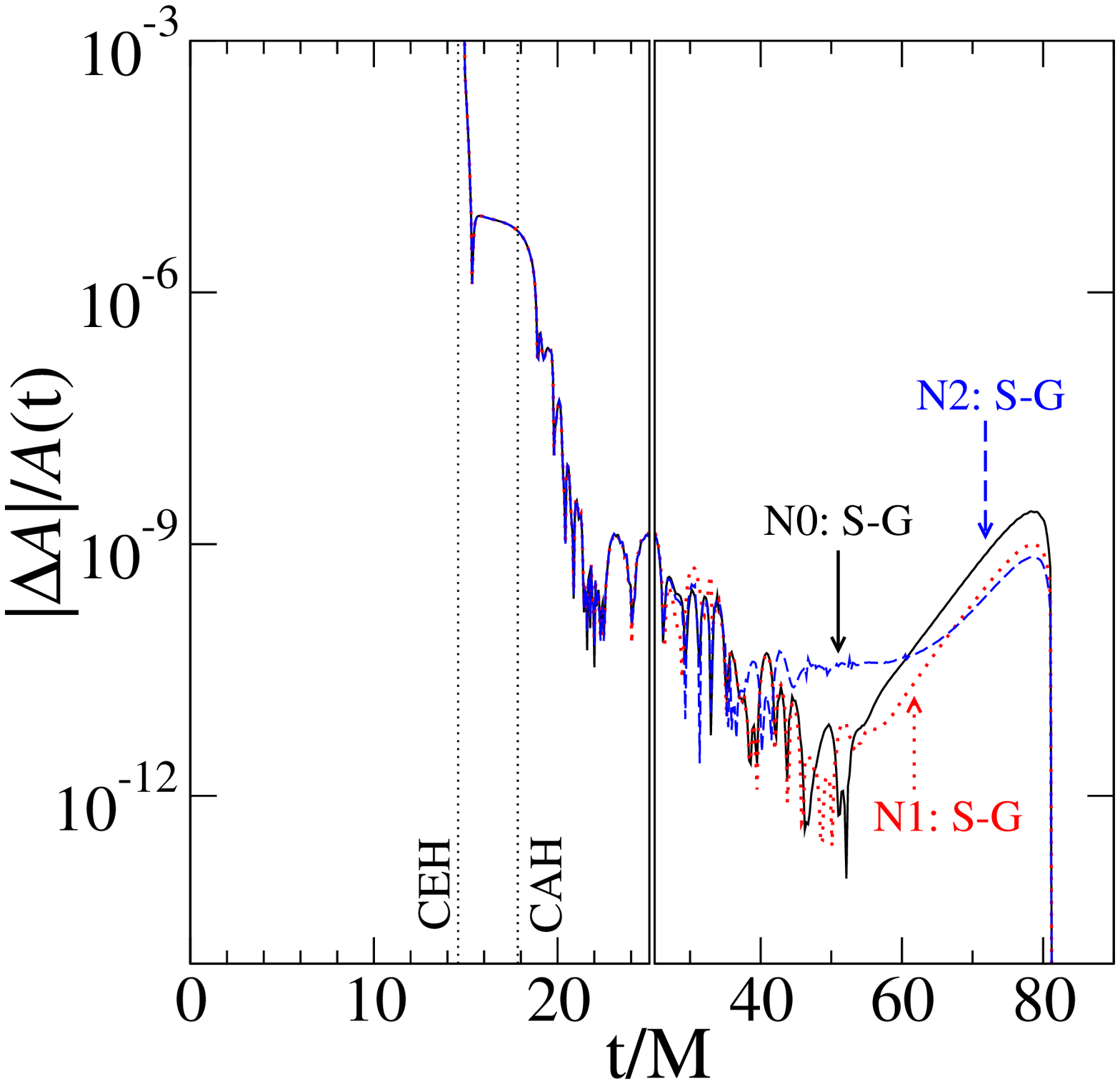}
\caption{Error estimates for the {\bf surface} and {\bf geodesic}
  methods, with surface resolution $L=47$.  The left panels show the
  root-mean-square pointwise deviation between the different runs,
  whereas the right panels show the differences in the surface area.
  The lines labelled ``G:N\#--N\#'' (``S:N\#--N\#'') in the upper
  panels show the difference between the geodesic method (surface
  method) when applied to merger simulations of different resolution
  N.  The lines labelled ``N\#:S--G'' in the lower panels show the
  differences between the surface and geodesic methods for a given N
  (where N0, N1, and N2 are resolutions of the merger simulation).
  Note that the time scale of all plots change at $t=25M$.
\label{fig:EHSEHGDiffs}}
\end{figure}

While the surface method presently cannot evolve through merger, it
nevertheless yields valuable consistency checks with the geodesic
method during the ringdown phase.  Figure~\ref{fig:EHSEHGDiffs}
presents such a comparison between the two methods and examines the
effect of varying the resolution of the underlying binary black hole
simulation.  The top panels show differences between the results of
the geodesic method applied to evolutions with different resolutions
(labelled ``G:N\#--N\#'').  As the underlying resolution is increased,
the differences become smaller.  Likewise, the lines labelled
``S:N\#--N\#'' show the analogous differences when running the surface
method.  When the surface method works, $t\gtrsim 15M$, it is more
accurate than the geodesic method.  For times close to the formation
of the common event horizon, $t\lesssim 15M$, errors in the surface
method grow very rapidly and render our current implementation
essentially useless.  The bottom panels of Fig.~\ref{fig:EHSEHGDiffs}
show differences between surface and geodesic method at the same
resolution of the evolved data.  This difference decreases with
increasing $N$, as it should.  During ringdown, $t\gtrsim 15M$, the
difference is essentially equal to the error in the geodesic method;
for $t\lesssim 15M$ it is dominated by errors in the surface method.

The right panels in Figure~\ref{fig:EHSEHGDiffs} examine the surface
area $A(t)$.  No clear convergence is apparent for $t\gtrsim 20M$,
perhaps because the surface area of the event horizon can be
calculated with great accuracy even at low values of $N$.  Given the
lack of clear convergence, we shall take as our error estimate for the
post-merger area the square sum of the following three error measures:
a) the change in $A(t)$ between the geodesic method applied to the
head-on simulation at the two highest resolutions (i.e. ``G:N1--N2''),
b) the change in $A(t)$ between the geodesic and surface methods
(i.e. ``N2:S--G'') and finally, c) the change in $A(t)$ in the
geodesic method at $L=47$, $N2$ when doubling the timestep (from
$0.056M$ to $0.112M$; the effect of this is small and not shown in
Figure~\ref{fig:EHSEHGDiffs}).  This combined error estimate is
plotted in Figure~\ref{fig:AreaRingdown}.

\subsection{Quasinormal Modes during Ringdown}
\label{ss:QuasiNormalModes}

\begin{figure}
\centerline{ \includegraphics[width=0.62\textwidth]{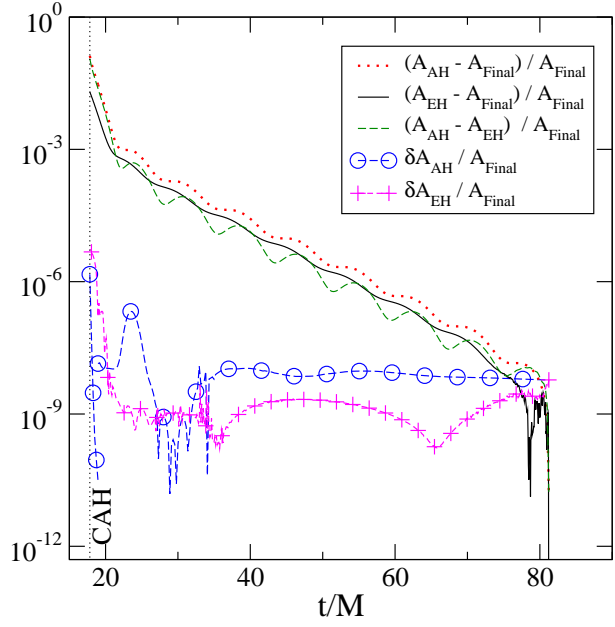} }
\caption{Surface area differences between the EH, AH, and the final
  area, normalized by the final area.  Also plotted are error
  estimates for $A_{\rm EH}(t)$ and $A_{\rm AH}(t)$. 
\label{fig:AreaRingdown}}
\end{figure}

After the merger, the distorted merged black hole rings down into a
stationary black hole.  During this phase, the area of the event
horizon, $A_{\rm EH}$ will approach its final value $A_{\rm Final}$,
and one expects that the apparent horizon approaches the event
horizon.  This is explored in Figure~\ref{fig:AreaRingdown}.  This
plot also contains the error estimates obtained from
Figures~\ref{fig:EHSEHGConvergence} and~\ref{fig:EHSEHGDiffs}.
Figure~\ref{fig:AreaRingdown} shows that the areas of the common AH
and EH differ by about 10\% when the common AH first appears, though
this difference drops to 0.1\% within about $3M$.  After this rapid
initial drop, the ringdown is clearly apparent.  The area of both EH
and AH approach their final area exponentially, and this approach is
resolved through about five orders of magnitude.  A least-squares fit
of $\log\left[\Afinal-A_{\rm EH}(t)\right]$ to the function $C - \lambda_{\rm
  obs} t$ for $30M\lesssim t \lesssim 70M$, yields $\lambda_{\rm
  obs}=0.181\Mfinal^{-1}$.  There are furthermore periodic features
visible in the EH and AH areas, with seven periods clearly
distinguishable.  The period of oscillation is found to be $\tau_{\rm
  osc}=8.00 M$, therefore $\omega_{\rm obs} = 0.745\Mfinal^{-1}$.

Decay rate $\lambda_{\rm obs}$ and frequency $\omega_{\rm obs}$ can be
related to quasi-normal modes of a Schwarzschild black hole as
follows: The quasinormal mode parameters of a perturbed black hole are
typically defined with reference to oscillations in the metric fields,
which can be written as
\begin{equation}
\delta g_{\mu\nu} \propto e^{-\lambda t} sin(\omega t),
\end{equation}
where $\lambda$ is the decay coefficient and $\omega$ is the angular
frequency of the metric oscillation.  Therefore
\begin{equation}
\delta \dot g_{\mu\nu} \propto -\lambda e^{-\lambda t} sin(\omega t) + \omega
e^{-\lambda t} cos(\omega t).
\end{equation}
The energy flux through the horizon, and therefore the change of its mass is
$\dot M \propto |\delta \dot{g}_{\mu\nu}|^2$, so we have
\begin{equation}
\frac{\dot A}{A} \propto \dot M \propto \frac{e^{-2\lambda t}}{2}[\lambda^2 + \omega^2 +
  (\omega^2-\lambda^2)cos(2\omega t) - \lambda \omega sin(2\omega t)].
\label{e:MassQuasinormalMode}
\end{equation} 
Thus the observed values $(\lambda_{\rm obs}, \omega_{\rm obs})$
should be {\em twice} the values $(\lambda, \omega)$ of a quasi-normal
mode.  Indeed, the lowest quasinormal mode of a perturbed
Schwarzschild black hole is the $\ell=2$, $n=0$ mode,
with~\cite{Kokkotas1999} $\lambda_{20}=0.08896\Mfinal^{-1}$ and
$\omega_{20}=0.37367\Mfinal^{-1}$.  Consistent
with~(\ref{e:MassQuasinormalMode}), we find that $\lambda_{\rm
  obs}-2\lambda_{20} = 0.003\Mfinal^{-1}$, and $\omega_{\rm obs} -
2\omega_{20} = 0.002\Mfinal^{-1}$.

\subsection{Treatment of Merger}
\label{ss:TreatmentOfMerger}

Before examining the merger phase in detail, we must develop tools to
analyse the topology change the event horizon undergoes during merger.
As seen in Figure~\ref{fig:MultipleCross-sections}, prior to merger,
the surface found by the event horizon finder is the union of the two
individual event horizons and the set of future generators of the
joint event horizon.  The event horizon itself consists of two
topological spheres.  At merger, $t=t_{\rm CEH}$, the topology of the
event horizon changes to a sphere.  For $t<t_{\rm CEH}$, generators of
the event horizon continuously enter the event horizon at the cusps on
the event horizons of the two approaching holes.  The geodesic method
traces geodesics perfectly fine through merger back to the start of
the head-on binary black-hole evolution, and the trajectories of the
geodesics are convergent as the resolution of the underlying evolution
is increased, see the left panels of Figure~\ref{fig:EHSEHGDiffs}.  In
this section, we address two questions relevant to analysing the
output of the geodesic method: First, when going toward earlier times,
some geodesics leave the event horizon; how does one decide whether a
given geodesic is still on the event horizon, or whether it is merely
a future generator of the event horizon?  Second, how can one compute
the area of the event horizon (i.e. not counting the area of the locus
of future generators)?

Let us first consider the area element $\sqrt{h}$ of the EH surface,
with $h$ given by (\ref{e:SurfaceAreaElement}), which
requires derivatives $\partial_u$,
$\partial_v$ along the surface, thus connecting neighbouring geodesics.
Because we place the geodesics at a $(u,v)$ grid consistent with
spherical harmonic basis functions, we can use spectral
differentiation to compute these derivatives (and have done so, up to
this point in the paper).  Convergence of this spectral expansion,
however, becomes increasingly slow for $t\lesssim t_{\rm CEH}$, and
therefore, we compute henceforth the derivatives $\partial_ur^i$ and
$\partial_v r^i$ with second order finite difference stencils.

\begin{figure}
\centerline{\includegraphics[width=.95\textwidth]{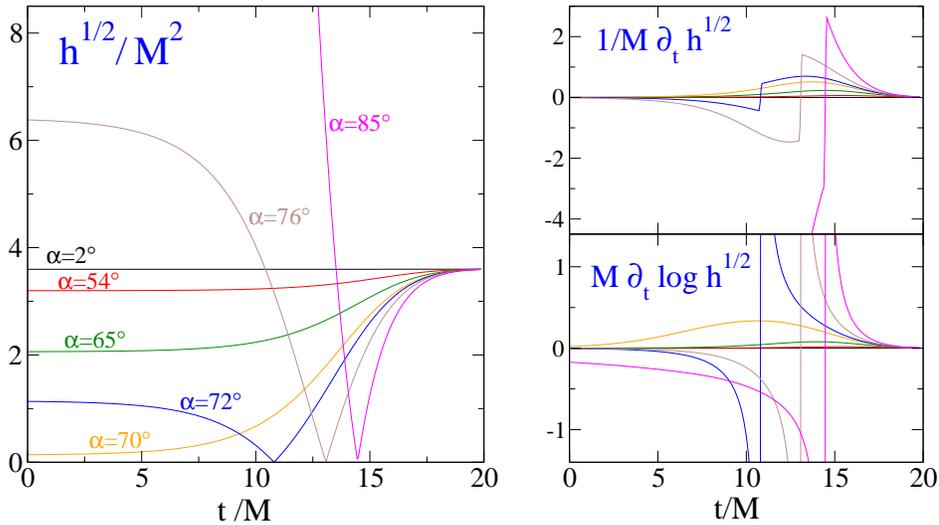}}
\caption{\label{fig:HeadonAreaElement}{\bf Left panel:} Area element
  $\sqrt{h}$ along a few representative geodesics during the head-on
  merger.  Each geodesic is labelled by the angle $\alpha$ between the
  initial location of the geodesic (at $\tend$) and the axis of
  symmetry.  Three types of behaviour are apparent: Geodesics entering
  the horizon from ${\cal I}^-$ ($\alpha=85^\circ$); geodesics
  entering the horizon from an area in the vicinity of the individual
  event horizons before merger ($\alpha=72^\circ$ or $76^\circ$), and
  geodesics remaining on the horizon throughout.  The right panels
  show the time derivative of $\sqrt{h}$, highlighting the clear
  signature when a geodesic enters the horizon.
}
\end{figure}

Figure~\ref{fig:HeadonAreaElement} plots the area element $\sqrt{h}$
as a function of time for a few representative geodesics.  This figure
was obtained from our highest resolution run using 20,000 geodesics.
To reduce CPU cost, these geodesics were initialized at $t=19.8 M$
from the $L=48$ run of the surface method.  For some geodesics in
Figure~\ref{fig:HeadonAreaElement}, $\sqrt{h}$ approaches zero at a
certain time.  This feature can be used to determine whether a given
geodesic is still on the horizon: We first note that the change of
area element along a given null geodesic (i.e. for fixed $u,v$) is
proportional to the expansion of this particular geodesic:
\begin{equation}\label{eq:hdot-propto-theta}
\partial_t \log\sqrt h=\frac{\partial_t(\sqrt{h})}{\sqrt{h}} \propto
\theta.
\end{equation}
The constant of proportionality depends on the parameterization of the
null geodesic.  Note that by Raychaudhuri's equation, the expansion of
a generator of the event horizon must be non-negative, $\theta\geq 0$.
Figure~\ref{fig:HeadonAreaElement} shows the area element as a
function of time for a few representative geodesics.  

At late time $t=\tend$ where the final black hole has settled down, we
start with geodesics on the apparent horizon, which will be very close
to the event horizon.  Therefore, we assume that at $\tend$ all
tracked geodesics are generators of the event horizon.  Consistently
with this assumption, Figure~\ref{fig:HeadonAreaElement} shows that
$\partial_t\log\sqrt h$ starts out very close to zero, and increases
as we approach the dynamical time region around merger.  If a
generator remains on the event horizon, $\partial_t\log\sqrt{h}$ will
eventually decrease again and approach zero at very early times before
the merger.  Generators leaving the event horizon must do so at points
where generators cross, according to a theorem by
Penrose~\cite{BattelleRencontres,MTW}.  For the head-on merger, at
such a point nearby geodesics cross and pass through each other.  Just
after the geodesic enters the horizon, the horizon generators diverge
from each other and their expansion is {\em positive} (and so is
$\partial_t\log\sqrt{h}$).  Just before the caustic points, nearby
future generators of the event horizon converge toward the caustic
point with {\em negative} expansion.  In fact, at the caustic,
$\partial_t\sqrt h$ changes sign discontinuously, as can be seen in
Figure~\ref{fig:HeadonAreaElement}.

 Therefore, the largest time at which the expansion of a geodesic
 passes through zero will be the time it joins the event horizon,
\begin{equation}\label{eq:GeodesicOff}
\partial_t\log\sqrt{h} \;\;
\Bigg\{\begin{array}{ll}
\;\leq\; 0, & t=t_{\rm join},\\ 
\;>\; 0,& t>t_{\rm join}.
\end{array}
\end{equation}
In practice, we keep track of (\ref{eq:GeodesicOff}) with a mask
function $f_M(u,v)$, which is initially identical to unity.  As we
evolve backward in time, we evaluate $\partial_t\log\sqrt{h}$ at each
time step, and if it drops below some tolerance $-\mbox{\tt tol}$ for
a point $(u_o,v_0)$ we set $f_M(u_0,v_0)=0$ for that geodesic.  The
tolerance {\tt tol} is necessary to avoid misidentifications due to
numerical truncation error at very early or late times, where
$\partial_t\log\sqrt{h}\to 0$ for event horizon generators.  Because
$\partial_t\log\sqrt{h}$ changes so rapidly at a caustic, the precise
value for {\rm tol} is not very important; we use ${\tt tol}=10^{-3}$.

For generic situations, generators can also leave the EH at points
where finitely separated generators cross (a ``cross--over point'' in
the language of Husa \& Winicour~\cite{Husa-Winicour:1999}).  At such
points, $\sqrt{h}$ remains positive, and criterion
(\ref{eq:GeodesicOff}) reduces to a necessary but not sufficient
condition that a generator has left the horizon, i.e. $t_{\rm join}$
from (\ref{eq:GeodesicOff}) will be a lower bound for the actual time
when a particular geodesic leaves the horizon.  Cross-over points
could be diagnosed by monitoring the minimal distance between every
pair of followed geodesics, and we shall discuss this point in more
detail in a future publication.

\begin{figure}
\centerline{\includegraphics[width=0.62\textwidth]{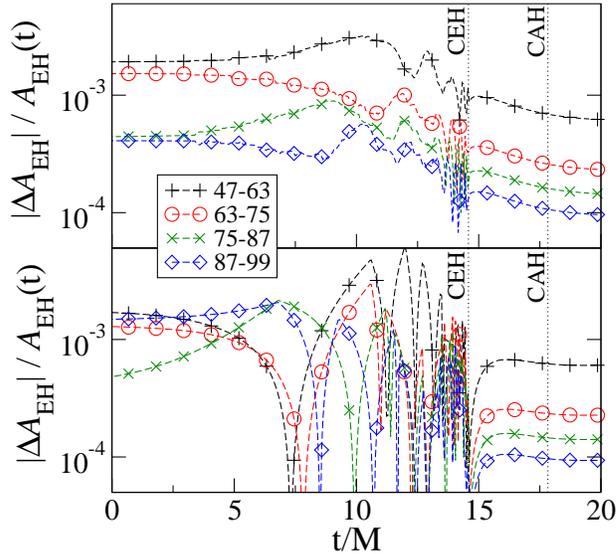}}
\caption{Convergence of the surface area of the event horizon during
  merger.  The lower plot shows results for placement of the geodesic
  pole parallel to the axis of symmetry (i.e. consistent with
  axisymmetry), the upper plot has a geodesic axis perpendicular to
  the axis of symmetry of the merger.  In both cases, geodesics are
  tracked using the geodesic method; derivatives for $\sqrt{h}$
  (cf. (\ref{e:SurfaceAreaElement})) are computed with
  finite-differences; geodesics are removed from the event horizon
  based on (\ref{eq:GeodesicOff}).  Lines are the difference between
  each resolution and the next highest.
\label{fig:EHG44MaskArea}}
\end{figure}

The area of the event horizon (consisting of the two disjoint
components for $t<t_{\rm CEH}$) is found by multiplying
$\sqrt{h}$ by the mask function $f_M$, and integrating:
\begin{equation}\label{eq:MaskedArea}
  A_{\rm EH}=\int f_M(u,v) \sqrt{h(u,v)}\;\sin u\;du\;dv.
\end{equation}
For $t<t_{\rm CEH}$, there are two major sources of error in this
integral: First, each geodesic can either be on or off the horizon.
When $f_M$ changes discontinuously from 1 to 0 for a geodesic, the
area of the event horizon will change discontinuously.  Note that this
will occur at different times for different resolutions.  The severity
of this effect will depend on how many geodesics enter the horizon
simultaneously, as illustrated by Figure~\ref{fig:EHG44MaskArea}.
This figure shows the convergence of the event horizon area with
increasing number of geodesics, and for two distinct orientations of
the geodesics.  In either case, the geodesics are initialized at
$t=19.86 M$ from the $L=47$ surface method determining the event
horizon during ringdown, and in either case the geodesics are placed
on a rectangular $(u,v)$ grid as detailed in
Section~\ref{ss:SpectralSurfaceImplementation}.  In the lower panel of
Figure~\ref{fig:EHG44MaskArea}, the geodesics are oriented {\em
  respecting} the axisymmetry (i.e. the $u=0$ polar axis is aligned
with the axis of symmetry), whereas in the upper panel the $u=0$ axis
is perpendicular to the axis of symmetry.  The lower panel of
Figure~\ref{fig:EHG44MaskArea}, with geodesics respecting the
symmetry, shows much larger variations in the area as the resolution
is increased.  This arises because due to the symmetry, a full {\em
  ring} of geodesics leaves simultaneously, thus amplifying the
discontinuity of $A_{\rm EH}(t)$.  For perpendicular orientation of
the geodesics, individual geodesics leave the horizon, resulting in
smaller jumps; this is the configuration we will use in the next
section to examine the physics of the black hole merger.

The second source of error in the evaluation of (\ref{eq:MaskedArea})
arises because the integrand is not smooth once geodesics have left
the horizon.  For fixed $t<t_{\rm CEH}$, $\sqrt{h}$ approaches zero
linearly toward the caustic; off the horizon, $f_M\sqrt{h}\equiv 0$ by
virtue of the mask function, so overall, the integrand is only
continuous, and we cannot expect exponential convergence of the
integral, despite using a Gauss-quadrature formula to evaluate
(\ref{eq:MaskedArea}).\footnote{For $t>t_{\rm CEH}$, $A_{\rm EH}$ in
  Figure~\ref{fig:EHG44MaskArea} is limited by the finite-difference
  derivatives used to compute $\sqrt{h}$.  Better accuracy can be
  obtained using spectral derivatives, as can be seen from the right
  panels of Figure~\ref{fig:EHSEHGDiffs}.  For the analysis of the
  merger below, this difference is invisible.}

\subsection{Analysis of Merger Phase}
\label{ss:MergerPhase}

\begin{figure}
\centerline{ \includegraphics[width=0.75\textwidth]{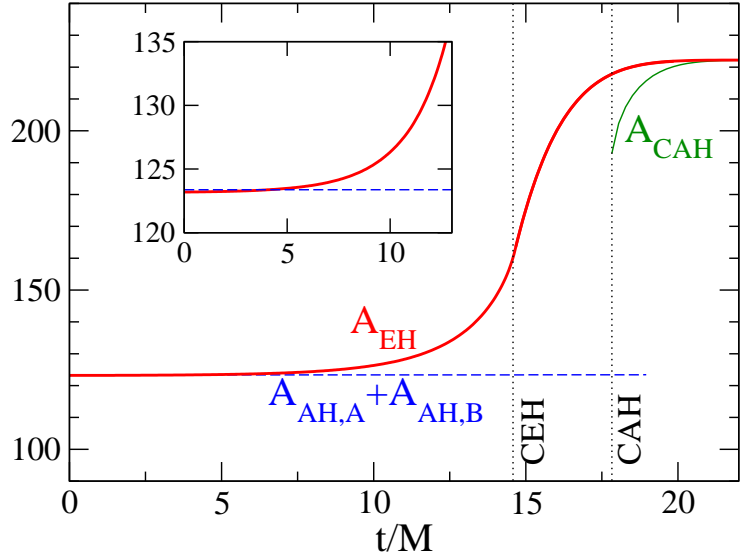} }
\caption{Area of the event horizon and of the apparent horizons
  before merger and during merger.  The vertical dotted lines indicate
  formation of a common event horizon and appearance of a common
  apparent horizon; the inset shows an enlargement for early time.
\label{fig:AHEHAreaHeadOn}}
\end{figure}

When evolving geodesics backward, we find that the first geodesic
leaves the horizon at $t_{\rm CEH}=14.58M$, the time of merger.
However, it should be noted that the point at which an observer sees
the EH change topology is not invariant because the curve traced by
the cusps of the two black holes is spacelike~\cite{Shapiro1995}.
Figure~\ref{fig:AHEHAreaHeadOn} shows the surface area of the EH and
the common and individual AHs during the merger phase.  The common
apparent horizon forms at $t_{\rm CAH}=17.8M$, and we track the
individual apparent horizons up to $t=18.8M$.  The area of the
individual apparent horizons is remarkably constant.  Up to formation
of the common event horizon, its fractional increase is less than
$10^{-5}$; up to common apparent horizon, its fractional increase is
$5\cdot 10^{-5}$, and even when we stop tracking the inner horizons,
their area has increased by only $1.6\cdot 10^{-4}$.  In contrast,
$A_{\rm EH}$ varies significantly more and at significantly earlier
times, as can be seen from the inset.

\begin{figure}
\centerline{ \includegraphics[width=0.75\textwidth]{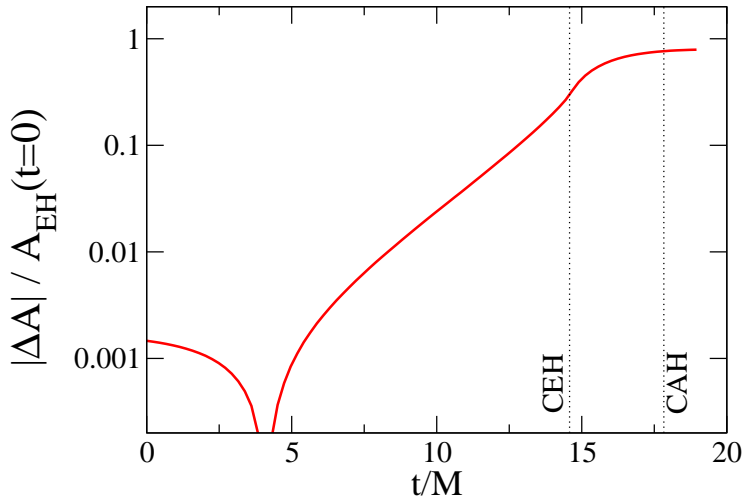} }
\caption{Difference between event-horizon area and the sum of the
  individual apparent horizon areas.  The vertical dotted lines
  indicate formation of common event horizon and appearance of a
  common apparent horizon.
\label{fig:AreaDivergence}}
\end{figure}

To examine the relation between individual apparent horizons and event
horizons, we plot in Figure~\ref{fig:AreaDivergence} the difference
$\Delta A\equiv A_{\rm EH}-(A_{\rm AH,A}+A_{\rm AH,B})$.  For times
$6\lesssim t/M \lesssim 14$, $\Delta A$ grows exponentially with an
e-folding time of $1.95M$.  This e-folding time is within a few
percent of the surface gravity of a black hole with the {\em initial
  mass} of the black holes in the head-on simulation.  This confirms
that as geodesics are integrated backwards in time, the individual
components of the event horizon approach the individual apparent
horizons with the expected rate.  If our code were free from all
numerical errors, the curve in Figure~\ref{fig:AreaDivergence} would
continue to decrease exponentially as one proceeds backwards in time.
Instead, this curve saturates at $\Delta A/A \approx 0.1\%$ at $t=0$,
and in addition, a feature in $\Delta A$ appears at $t\approx 5M$
because the EH area falls below $A_{\rm AH,A}+A_{\rm AH,B}$ and
therefore $\Delta A$ changes sign. These effects are due to numerical
errors, particulary finite-difference errors in the computation of
$\sqrt{h}$ and the use of a finite number of geodesics.  If one wishes
to achieve much better than $0.1\%$ accuracy of the event horizon
surface area at very early times when the two holes are widely
separated, the EH must be split into two individual surfaces to be
evolved separately.

\section{Conclusion}
\label{s:Conclusion}

This paper examines three different methods for locating event
horizons in dynamical black hole spacetimes, the geodesic method, the
surface method and the level-set method.  All three methods rely on
the principle that outgoing geodesics exponentially approach the event
horizon when followed backward in time.  We implement both the
geodesic and surface methods, the latter one implemented without the
assumption of axisymmetry as done in earlier
work~\cite{Libson96}. Overall, we find that the geodesic method is
more robust, with the capability to accurately follow highly spinning
black holes (tested up to $a/M=0.99$), as well as the merger of two
black holes.  For the head-on merger, we find that the surface-area
element $\sqrt{h}$ of the geodesic congruence is an excellent
diagnostic of whether and when a geodesic joins the event horizon,
cf. (\ref{eq:GeodesicOff}).  In more generic situations, this
criterion might have to be amended by a second test for crossing of
geodesics that are initially (at $t=\tend$) separated by a finite
amount.  Errors due to tangential drift of the geodesics---as
explained in~\cite{Libson96}---are not apparent in our simulations.
The observed good properties of the geodesic method might be related
to the improvements in accuracy of the spacetime metric since the
early tests~\cite{Libson96}, as well as the ability to interpolate the
metric spectrally to the geodesic locations.  Because each geodesic is
evolved independently, the geodesic method parallelizes trivially.
Tracking of the cusp of the disjoint components of the event horizon
before merger, as well as computation of $A_{\rm EH}$ is currently not
highly accurate, as comparatively few geodesics cover the region close
to the cusps.  Our current scheme switches from the surface method to
a large number of geodesics some time after merger where the surface
method is still very accurate, say at $t_0$.  We plan on adaptively
placing additional geodesics at $t_0$ based on where cusps occur.
  
The surface method is less robust and exhibits a long-term instability
when applied to Kerr black holes with spins $a/M\lesssim 0.6$, and
rapid blow-up for larger spins.  Nevertheless during the ringdown
phase $t>t_{\rm CEH}$ of the head-on merger, the surface method
locates the event horizon with comparable accuracy to the geodesic
method and provides an important independent test of the geodesic
method.  However, when the surface being tracked self-intersects in a
caustic point, our current method for defining the normal breaks down
because $\partial r^i/\partial v = 0$
in~(\ref{e:SurfaceSpatialDerivatives})-(\ref{e:Signoflambda}), and
thus our current implementation of the surface method fails.

The level-set method, finally, is not implemented in this paper.  It
requires boundary conditions for the level-set function $f$;
furthermore $f$ can become singular during a black hole merger. Both
reasons made it unduly difficult to implement this method in our
spectral code.  In conclusion, we find that the geodesic method, the
oldest of the three methods considered, to be the most accurate and
useful in our tests.

Turning our attention to applications of the event horizon finders,
Figure~\ref{fig:KerrSurfaceGravity} presents a new quantitative test of
event horizon finders: When finding the EH of a Kerr black hole
starting away from the true horizon, does the tracked null surface
approach the true event horizon with the correct rate, namely the
surface gravity $\gH$?  Table~\ref{tab:SurfaceGravity} confirms this
for the geodesic method.  For the head-on  merger, both
geodesic and surface method perform admirably during the ringdown
phase, where we are able to clearly observe the quasinormal ringing of
the single merged black hole.  For both the event and apparent
horizons, the frequency and damping time of the ringing matches the
$(\ell=2,n=0)$ mode of the Schwarzschild quasinormal ringing spectrum
to within 2\% for the decay rate and 0.3\% for the frequency.

Furthermore, we find that the apparent horizons provide an excellent
approximation to the event horizon for the head-on merger very early
before the merger, and very late after the ringdown.  Thus, while in
principle the apparent horizon is slice-dependent and there is no
guarantee that it should coincide with the event horizon, in practice
no such behaviour is found.

\begin{figure}
\centerline{\includegraphics[width=0.85\textwidth]{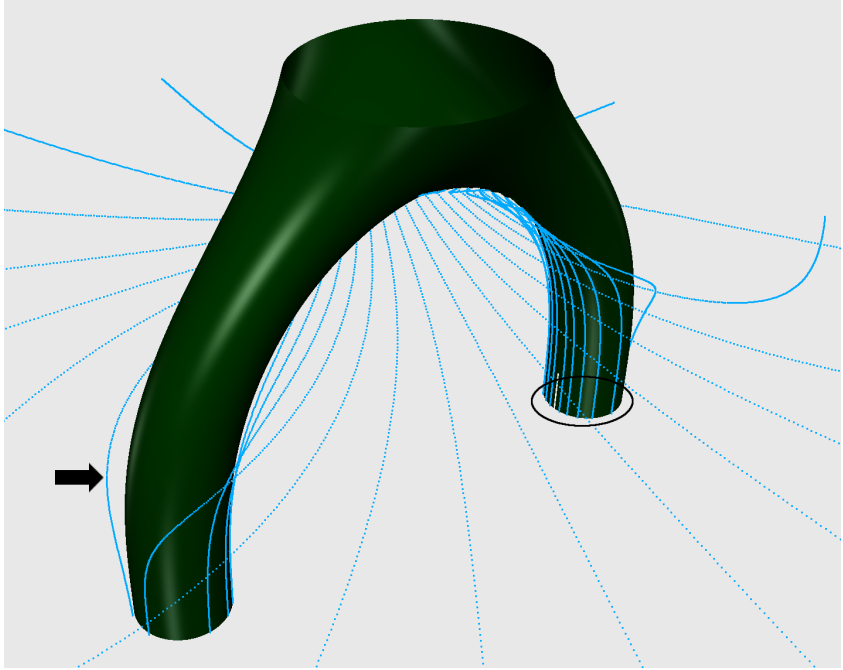}}
\caption{\label{fig:Spacetime} Spacetime diagram of the head-on
  merger.  The pale lines denote geodesics that will join the event
  horizon.  Some of these geodesics come from past null infinity, but
  others come from a region close to the individual event horizons
  (cf. the arrow and the circled geodesics on the far black hole).  }
\end{figure}

Finally, perhaps surprisingly, for the head-on binary black hole
merger only {\em some} of the future null generators of the horizon
start at past null infinity.  A significant fraction of the
generators rather start close to the individual event horizons of the
black holes before merger.  This can be seen in the spacetime diagram
in Figure~\ref{fig:Spacetime}, most clearly for the geodesic pointed
to with an arrow.  These geodesics begin to diverge from the
individual event horizon as the second black hole approaches.  The
increased gravity of both black holes causes such geodesics then to
``turn around'' and join the event horizon at the seam of the pair of
pants.

In the future we plan to study event horizons in the more diverse
black hole scenarios currently being simulated: mergers of inspiraling
black holes; spinning and/or non-equal mass binary black holes, as
well as black hole Neutron star mergers.

\ack We gratefully acknowledge useful discussions with Lee Lindblom,
Peter Diener, Saul Teukolsky and Kip Thorne.  We particularly thank
Kip Thorne for pointing out the relationship between the surface
gravity and the rate of divergence of geodesics from the EH.  This
work was supported by grants from the Sherman Fairchild Foundation and
the Brinson Foundation, and by NSF grants PHY-0601459, PHY-0652995,
DMS-0553302 and NASA grant NNG05GG52G.

\appendix
\section{Points in the Surface Method follow geodesics}
\label{App:SurfaceGeodesicProof}

Consider a 2-dimensional family of null geodesics, $q^\mu(t, u, v)$,
where $u,v$ label different geodesics. Assume the parameter $t$ along
the geodesic coincides with the coordinate time of the underlying
black hole simulation, i.e. $q^0(t, u, v)=t$.  This family of
geodesics traces out a three-dimensional null surface ${\cal N}$,
parameterized by coordinates $t,u,v$: $q^\mu(t, u, v)$, where $t$ is
the parameter along each null curve, and $u,v$ are the parameters
relating each null curve to nearby null curves.  In this
parameterization, we can write the outgoing null normal $\ell^\mu =
\left.\partial q^\mu/\partial t\right|_{u,v}$, i.e. a coordinate
derivative $\vec{\ell}=\partial_t$ within the $(t,u,v)$ coordinates of
${\cal N}$.  Displacement vectors that relate each null curve to its
neighbours are given by $\vec m = \partial/\partial u$, $\vec n =
\partial/\partial v$.  Since coordinate derivatives commute, we have
\begin{equation}
\ell^\mu \nabla_\mu m^\nu = m^\mu \nabla_\mu \ell^\nu.
\label{e:Commutation_l_and_m}
\end{equation}
Let us consider the rate of change of the inner product $\ell^\mu
m_\mu$ as we change the time $t$ along a geodesic (i.e. for fixed $u$
and $v$):
\begin{eqnarray}
\partial_t (\ell^\mu m_\mu) &=&
\ell^\nu \nabla_\nu (\ell^\mu m_\mu)
=m_\mu\ell^\nu\nabla_\nu(\ell^\mu) +
\ell_\mu\ell^\nu\nabla_\nu(m^\mu). \label{e:time_change_of_l_and_m}
\end{eqnarray}
From (\ref{e:Commutation_l_and_m}), the second term of
(\ref{e:time_change_of_l_and_m}) vanishes,
\begin{equation}
\ell_\mu\ell^\nu\nabla_\nu(m^\mu)=\ell_\mu m^\nu\nabla_\nu(\ell^\mu)
=\frac{1}{2}m^\nu\nabla_\nu(\ell^\mu\ell_\mu)=0.
\end{equation}
Substituting the formula for parallel transport of $\ell^\mu$ along
the geodesics, $\ell^\nu\nabla_\nu\ell^\mu=\kappa\, \ell^\mu$ (with
$\kappa=0$ if $t$ is affine), (\ref{e:time_change_of_l_and_m}) finally
becomes
\begin{equation}
\partial_t (\ell^\mu m_\mu) = m_\mu\ell^\nu\nabla_\nu(\ell^\mu) =
\kappa\, m_\mu \ell^\mu \label{e:rateofchangeofelldotm}.
\end{equation}
A similar calculation results in $\partial_t(\ell^\mu n_\nu)=\kappa\,
\ell^\mu n_\mu$.

So far, this appendix only discusses the geodesic method.  We now use
the results just obtained to show that surface and geodesic methods
will construct the same null surface ${\cal N}$.  Both methods start
with the same two-dimensional surface at some late time $t_0$, and the
tangent $\dot{q}^\mu(t_0,u,v)$ to the geodesics at $t_0$ is chosen to
be normal to the 2-surface.  Therefore, at $t_0$,
$\ell^\mu=\dot{q}^\mu$, and the surfaces resulting from evolving both
the geodesic and surface methods will coincide at times
infinitesimally near $t_0$.  Because $\ell^\mu m_\mu=\ell^\mu
n_\mu=0$ initially, (\ref{e:rateofchangeofelldotm}) implies that
$\ell^\mu m_\mu=\ell^\mu n_\mu = 0$ at all other times.  Thus, the
tangent to the geodesics always remains orthogonal to the surface
described by the positions of all the geodesics at a given time $t$.
Since $\dot{q}^\mu$ is normal to that surface, null, outgoing, and has
$\dot q^0 = 1$, it is identical at all times to $\ell^\mu$ as
constructed by the surface method.  Therefore, we see that the
surfaces obtained by the geodesic and surface methods agree, and both
techniques trace out the same $\cal N$ given the same initial
conditions.

\section{Proof of Surface Gravity conjecture}
\label{App:SurfaceGravity}

We consider a null geodesic $q^\mu(t)$ that asymptotes to a horizon
generator $q^\mu_H(t)$ for $t\to-\infty$, i.e.
\begin{equation}\label{eq:q+deltaq}
q^\mu(t) = q^\mu_H(t) + \delta q^\mu(t)
\end{equation}
with $\delta q^\mu(t)\to 0$ as $t\to-\infty$.  In the discussion of
Figure~\ref{fig:KerrSurfaceGravity} we have asserted that
\begin{equation}\label{eq:deltaq-soln}
  \delta q^\mu(t) \propto e^{\gH t},
\end{equation}
where $\gH$ is the surface gravity of the black hole, and where the
coordinates $x^\mu$ are Kerr-Schild coordinates,
cf. (\ref{eq:Kerr-Schild})--(\ref{eq:KerrRsquaredDef}).  To confirm
this assertion, one can substitute (\ref{eq:q+deltaq}) into the
geodesic equation and expand to linear order in $\delta q^\mu$ (where
we assume that $\delta q^\mu$, $\delta\dot q^\mu$, and $\delta\ddot
q^\mu$ are of the same order).  One then needs to show that the
resulting linear equation indeed has the solution
(\ref{eq:deltaq-soln}).

The linearization of the geodesic equation is most easily performed in
adopted coordinates.
We have performed the analysis in ``rotating spheroidal Kerr-Schild
coordinates'' $x^{\mu'}=(t,r_{\rm BL},\theta,\phi)$, related to the
standard Kerr-Schild coordinates of
(\ref{eq:Kerr-Schild})--(\ref{eq:KerrRsquaredDef}) by the coordinate
transformation
\begin{eqnarray}\label{eq:Spheroidal-x}
x&=&\sqrt{r_{\rm BL}^2+a^2}\sin\theta\cos\left(\phi+\Omega_H
t\right),\\ y&=&\sqrt{r_{\rm
    BL}^2+a^2}\sin\theta\sin\left(\phi+\Omega_H t\right),\\
\label{eq:Spheroidal-z}
z&=& r_{\rm BL}\cos\theta.
\end{eqnarray}
The time $t$ is not transformed.  Horizon generators have the form
$q^{\mu'}=[t, r_+, \theta_0, \phi_0]$, with $r_+=M+\sqrt{M^2-a^2}$ and
$\theta_0, \phi_0$ constants, i.e. $\dot q_H^{\mu'}\propto[1,0,0,0]$.
In these coordinates, we have considered the geodesic equation in
affine parameterization, (\ref{e:geodesic}) and have indeed confirmed
\begin{equation}\label{eq:deltaq-soln2}
\delta q^{\mu'}\propto e^{\gH t}
\end{equation}
to leading order in $\delta q^{\mu'}$.  Exponential divergence from a
horizon generator---as in (\ref{eq:deltaq-soln2})---is a property
present in a quite general class of coordinate systems.  For instance,
consider the coordinate transformation
\begin{equation}
t' = t + f(x^i), \qquad x^{i'} = x^{i'}(x^i),
\end{equation}
where the Jacobian $\partial x^{i'}/\partial x^i$ and its inverse are
finite in a neighborhood of the horizon. In this case, $\delta
q^{\mu'}$ and $\delta q^{\mu}$ are related merely by a multiplication
by the Jacobian, so the exponential behavior $e^{\gH t}$ is the same
in both coordinate systems.  The coordinate transformation
(\ref{eq:Spheroidal-x})--(\ref{eq:Spheroidal-z}) falls into this
class, and therefore (\ref{eq:deltaq-soln2}) implies
(\ref{eq:deltaq-soln}).

\section*{References}

\end{document}